\documentclass[acmsmall,screen,nonacm]{acmart}
\AtBeginDocument{%
  \providecommand\BibTeX{{%
    \normalfont B\kern-0.5em{\scshape i\kern-0.25em b}\kern-0.8em\TeX}}}

\usepackage{amsthm}
\usepackage{amsmath}
\usepackage{natbib}
\usepackage{bm, amsfonts,bbold}
\usepackage{subcaption}
\usepackage{booktabs}

\usepackage{enumitem}
\usepackage{tikz}
\usepackage{ifthen}
\usepackage{color}

\usepackage[title]{appendix}

\DeclareRobustCommand\circled[1]{\tikz[baseline=(char.base)]{
            \node[shape=circle,draw,inner sep=2pt] (char) {#1};}}

\newcommand{\diamonddiamond}{%
\begin{tikzpicture}[x=0.75pt,y=0.75pt,yscale=-1,xscale=1]
\draw   (78.5,188.44) -- (82.36,191.83) -- (78.5,195.22) -- (74.64,191.83) -- cycle ;
\draw   (78.5,185.81) -- (85.5,191.81) -- (78.5,197.81) -- (71.5,191.81) -- cycle ;
\end{tikzpicture}
}

\newcommand{\ev}[1]{\mathrm{F}_{#1}}
\newcommand{\glob}[1]{\mathrm{G}_{#1}}

\newcommand{\somewhere}[2]{\diamonddiamond_{#1}^{#2}}

\newcommand{\reach}[2]{\mathcal{R}_{#1}^{#2}}
\newcommand{\escape}[2]{\mathcal{E}_{#1}^{#2}}

\newboolean{showcomments}
\setboolean{showcomments}{true} % toggle to show or hide comments
\ifthenelse{\boolean{showcomments}}
{\newcommand{\nb}[3]{
  \fcolorbox{black}{#3}{\bfseries\sffamily\scriptsize#1}
  {\sf\small\textit{#2}$\blacktriangleleft$}
 }
 
}
{\newcommand{\nb}[3]{}
 
}

\begin{document}

\title{Bayesian Machine Learning meets Formal Methods: An application to spatio-temporal data}

%\runtitle{Bayesian Machine Learning meets Formal Methods}
%\runtitle{Bayesian formal predictive model assessment}

%% Of note is the shared affiliation of the first two authors and the
\author{Laura Vana-G\"ur}
\email{laura.vana.guer@tuwien.ac.at}
\orcid{0000-0002-9613-7604}
\affiliation{%
  %\institution{Institute of Statistics and Mathematical Methods in Economics}
  \institution{TU Wien}
  \city{Vienna}
  \country{Austria}
}

\author{Ennio Visconti}
\orcid{0000-0002-1146-4850}
\authornote{Corresponding author}
\email{ennio.visconti@tuwien.ac.at}
\affiliation{%
  \institution{TU Wien}
  \city{Vienna}
  \country{Austria}}

\author{Laura Nenzi}
\orcid{0000-0003-2263-9342}
\affiliation{%
  \institution{University of Trieste}
  \city{Trieste}
  \country{Italy}
}

\author{Annalisa Cadonna}
\orcid{0000-0003-0360-7628}
\affiliation{%
 \institution{University of Klagenfurt}
 \city{Klagenfurt}
 \country{Austria}
}

\author{Gregor Kastner}
\email{gregor.kastner@aau.at}
\orcid{0000-0002-8237-8271}
\affiliation{%
  \institution{University of Klagenfurt}
  \city{Klagenfurt}
  \country{Austria}
}
%%
%% By default, the full list of authors will be used on the page
%% headers. Often, this list is too long and will overlap
\renewcommand{\shortauthors}{Vana, Visconti, Nenzi, Cadonna and Kastner}

\begin{abstract}
    We propose an interdisciplinary framework that combines Bayesian predictive inference, a well-established tool in Machine Learning, with Formal Methods rooted in the computer science community.
    Bayesian predictive inference allows for coherently incorporating uncertainty about unknown quantities by making use of methods or models that produce predictive distributions, which in turn inform decision problems. By formalizing these decision problems into properties with the help of spatio-temporal logic, we can formulate and predict how likely such properties are to be satisfied in the future at a certain location.  Moreover, we can leverage our methodology to evaluate and compare models directly on their ability to predict the satisfaction of application-driven properties.
    The approach is illustrated in an urban mobility application, where the crowdedness in the center of Milan is proxied by aggregated mobile phone traffic data. We specify several desirable spatio-temporal properties related to city crowdedness such as a fault-tolerant network or the reachability of hospitals. After verifying these properties on draws from the posterior predictive distributions, we compare several spatio-temporal Bayesian models based on their overall and property-based predictive performance.
\end{abstract}

%\begin{keyword}[class=MSC]
%\kwd[Primary ]{62F15}
%\kwd[; secondary ]{62C10, 62P12}
%\end{keyword}




%% This command processes the author, affiliation and title
\maketitle
%%%%%%%%%%%%%%%%%%%%%%%%%%%%%%%%%%%%%%%%%%%%%%
\section{Introduction}

Formal verification methods have a long-standing tradition in the computer science community. They have historically emerged in the context of hardware and software systems to provide strong guarantees about the correctness of the analyzed implementation by verifying the satisfaction of complex logical properties. From deterministic systems, formal verification methods have found plenty of applications in stochastic systems, where the same set of inputs can correspond to multiple sets of random outputs. The traditional approach to formal verification of stochastic systems is probabilistic model checking \citep[introduced independently by][]{clarke, queille}. 
However, in the context of very large stochastic systems, numerical probabilistic model checking is practically infeasible \citep{younes_numerical_2006}, and new approaches have been developed, such as Statistical Model Checking (SMC)
\citep[see][for a recent survey on the area]{Legay2019}, which is a simulation-based version of probabilistic model checking. 
The idea of SMC is to calculate the probability of satisfaction of logical properties by Monte Carlo integration, that is, simulate many trajectories from the stochastic system, determine for each trajectory if the logical property is satisfied, and take the arithmetic mean over the number of simulations.  Given the parallelizable nature of SMC, the computational bottleneck becomes the simulation and verification of a complex logical property at each iteration. Here, formal methods come into play: complex properties are translated into logical formulae, which can then be automatically verified using efficient algorithms tailored to the type of logic employed.  The primary advantage of specifying properties as logic formulae comes from the efficient monitoring algorithms that are available to automatically check whether the specified properties are satisfied or not, and to which extent. Given its scalability, SMC has therefore become increasingly used in different application domains, especially related to biological systems and cyber-physical systems \citep[see][for an application of a continuous-time Markov chain model to model a bike sharing system]{nenzi2017qualitative}. 

%\textbf{1) SMC for complex parametric stochastic models with prior on %parameters
%}

SMC applications usually consider the model parameters to be fixed to specific values (e.g., the maximum likelihood estimates) and simulate from the stochastic system conditioned on the fixed parameter values, without taking into account how the uncertainty on the parameter naturally propagates to the satisfaction probability. While there are extensions to SMC to include prior information about the probability of the satisfaction of a property, such as Bayesian SMC \citep{zuliani2013bayesian}, the finite set of trajectories remains simulated from a model with fixed parameter values. 
On the opposite side, to deal with such uncertainty,   \cite{BortolussiMS16, BortolussiS18, BortolussiCCP23} consider directly the parameters as uncertain and calculate the probability of satisfaction for a set of possible parameter combinations. However, while the range of the parameters was determined based on prior knowledge, the values of the parameters are not estimated based on observed data. 
%In most real-life applications, stochastic systems depend on a set of parameters whose true values can be estimated from the observed data.

In this paper, we propose a Bayesian Machine Learning approach that naturally deals with uncertainty propagation, while simultaneously it allows to learn the value of the parameters from the data. Our proposed approach extends the classical approach to SMC to a Bayesian framework by performing verification and monitoring on trajectories of the  Bayesian predictive distribution 
drawn using the MCMC algorithm employed for model estimation. Bayesian predictive inference allows the coherently accounting for uncertainty about an unknown or future value of the random variable being modeled by providing the entire posterior predictive distribution. 
This novel framework provides a unifying approach to the modeling and statistical analysis of data that coherently accounts for the uncertainty of the specified properties. 

While we believe a Bayesian approach to uncertainty can enrich SMC, we also argue that the use of Formal Methods in the (Bayesian) Machine Learning community opens the door for interesting application-driven methods for prediction and model evaluation and comparison. 
In most applications, predictions are obtained from a (Bayesian) Machine Learning model
which are then compared to the observed data in order to assess the model performance ex-post 
(i.e., post-estimation). However, predictions obtained from a (Bayesian) Machine Learning model are not directly translated into a decision but rather transformed, compressed, and combined with further rules or requirements relevant to the decision problem at hand. As a simple example, consider an algorithmic trading scheme using a predictive model for stock returns, which will use the output of the predictive model together with the rule ``place a sell order if the 90\% quantile of the predictive stock return distribution exceeds 10\% three days in a row''. Or a traffic officer who will decide to divert traffic if the model predicts crowdedness to rise above a certain threshold along the city's main arteries. Such rules could also be employed to put monitoring systems in place for the predictive models 
(especially black-box ones); e.g., for the purpose of ensuring fairness in the predictions. 
Especially in high-dimensional, complex models, these requirements or properties relevant for decision-making are typically highly nonlinear functions of the random variables, and one is interested in their predictive distribution. Their verification ex-ante as well as their evaluation ex-post (as part of the posterior model checking and comparison exercise) could provide valuable insights to the modeler and decision-maker and could tailor the analysis to the concrete needs of the decision problem.
We claim that the focus should be placed on application- and decision-specific requirements or properties that the system being modeled should satisfy also when evaluating and comparing the performance of different Machine Learning models. 

%From a decision-theoretic point of view, the predictive performance of a model is typically defined
%in terms of a utility or scoring function that measures the
%quality of the predictive distribution of candidate models \citep{bernardo1994bayesian}. 
%In the Bayesian literature, posterior predictive model evaluation and checking  \citep[][]{rubin1984bayesianly} is employed to check the predictive performance of a Bayesian model on unseen data by qualitatively and quantitatively assessing how well the posterior predictive distributions produced by a model reflect existing data. A large body of literature has been concerned with how models should be evaluated and compared based on these predictive distributions, with various versions of predictive density scores being proposed \citep[see, e.g.][]{CORRADI2006197, geweke2010comparing, frazier2021lossbased}. 
% We argue that the score functions or rules employed in the model assessment should be specifically tailored for the application at hand, and the model assessment and comparison should take into account the process through which the prediction of future data with the model enters a decision. 
% We advocate in this paper for the formulation and verification of complex spatio-temporal properties as part of the Bayesian workflow in data analysis. 
We achieve this by leveraging an existing stream of literature in the computer science field of SMC and verification to approximate the posterior predictive probability of satisfaction of these properties, as well as a posterior predictive measure of property reliability or robustness. We then introduce the Bayesian predictive probability of satisfaction and posterior predictive robustness as quantities of interest and show how these measures can be used for comparing a collection of spatio-temporal Bayesian models. This property-related comparison 
can complement common predictive evaluation measures such as the log predictive density scores.
%However, such approaches are rather limited in the literature for spatio-temporal models, where commonly used measures based on, e.g., logarithmic scores, are chosen for their desirable mathematical properties, as a measure of similarity between the model predictions and the true values. 
%%
%Especially when a temporal component is present both in the data and in the modeling framework, 
%the posterior predictive distribution for different time intervals will inform a series of decision problems. Clearly, in the decision-making process, these future values are unknown ex-ante but will be observed ex-post, i.e., after the decision has been taken, so competing models can be compared based on evaluating the out-of-sample posterior predictive distribution given the future observed values. These predictive distributions can be obtained either in an exact fashion, by employing a (time-series) cross-validation where the model is re-estimated as new observations come in, or in an approximate fashion \citep[see e.g.,][]{vehtari2017,burkner2020approximate}. 

%In the computation of these methods, we rely on draws from a Bayesian predictive distribution, which are obtained through the Markov chain Monte Carlo (MCMC) algorithms employed for parameter estimation.

%\textbf{Application}

We demonstrate our novel approach with spatio-temporal areal data, where measurements are collected over time at various areal units, and a neighboring matrix allows calculating the distance between the different units. In particular, we consider an urban mobility application, given that urban population density dynamics are highly variable both in space and time. For such applications, building a Bayesian spatio-temporal model that accurately predicts future population dynamics, is of paramount importance to decision-makers in the context of urban planning (e.g., who must plan for resource allocation, divert traffic and increase mobile network capabilities temporarily) but has far-reaching implications related to the environment, economy, and health \citep{data4010008}. In particular, the latter link became even more evident in the context of the COVID-19 pandemic. 
%%
%While traditional data collection methods regarding population mobility, such as census or surveys, are rather infrequent, high-frequency mobile phone traffic data has served as a viable alternative, with mobile phones being almost universally used worldwide.
Analyzing mobile phone traffic data as a proxy for population mobility has been widely employed in the past years \citep[e.g.,][and references therein]{deville2014dynamic, peters2017dynamic, data4010008, bernini2019time}, with applications ranging from population density estimation in the absence of census data \citep{wardrop2018spatially} to traffic prediction \citep[e.g.,][]{iqbal2014development} and to modeling the spread of epidemics \citep[e.g.,][]{cinnamon2016evidence,bonato2020mobile}.
Given the high dimensionality of mobile phone data, only a few studies have focused on sophisticated (Bayesian) modeling tools in an urban planning context. \cite{modelac} build a spatio-temporal model with spatial clustering of the locations in Milan using data from an Italian telecommunications company; \cite{wang2021bayesian} conducted an empirical study in Shenzhen, China where they include the population statistics and indices for mixed-use to explore the spatial pattern of population fluctuation in a Bayesian model. 
% Time series clustering using a Poisson-Dirichlet prior has been proposed in \cite{nieto2014bayesian} who apply the model to share price data from the Mexican stock exchange.

For illustration purposes, we employ in this work open source data from the ``Telecom Italia Big Data Challenge'', which contains telecommunications activity aggregated over a fixed spatial grid of the city of Milan during the months of November and December 2013. 
Our results provide a deeper understanding of urban dynamics in Milan in terms of the best-performing model which identifies clusters of areas with similar temporal patterns and in terms of when and how well the formulated properties are satisfied. 
% Additionally, the results exemplify the value of combining Bayesian predictive modeling with formal verification methods in terms of insights to be gained. 

The paper is organized as follows: 
Section~\ref{sec:formalver} introduces the concept of formal verification for spatio-temporal data, together with new application-specific properties. 
%The spatio-temporal properties that are to be verified are introduced in Section~\ref{sec:properties}.
Section~\ref{sec:bayes} explains how the Bayesian predictive framework allows for the incorporation of uncertainty. Section~\ref{sec:property_comparison} proposes measures for predictive evaluation in terms of property satisfaction.
The empirical data and results are presented in Section~\ref{sec:results}.
Section~\ref{sec:conclusion} concludes and outlines directions for future work.

\section{Formal verification for areal spatio-temporal data} \label{sec:formalver}
\subsection{Introduction to STREL logic}
\label{sec:strel}
With the goal of incorporating application-specific properties into the Bayesian Machine Learning workflow, 
we introduce formal verification methods as a way to specify and verify such properties.
A formal verification method has the goal of checking whether a (stochastic) system satisfies
some properties or requirements, which are stated in some formal language. The last decade has seen a
great effort to develop logic-based specification languages and monitoring frameworks for spatio-temporal properties;
in our case, we consider STREL \citep{lmcs_8936} as the specification language of reference.
A spatio-temporal logic combines atomic propositions via a set of operators: the standard Boolean operators ($\vee$, $\neg$, $\rightarrow$,\ldots), temporal operators 
to specify the temporal evolution and spatial operators to reason about the space. Let us describe the language more in detail and how it can be applied to spatio-temporal data. 

In our application, we consider a grid of  $i = 1, \ldots, I$ areas, which covers the center of the city of Milan, where a measurement of crowdedness $y$ is collected over time at  $t = 1, \ldots, T$, where one time-point represents a time interval of 10 minutes.  The distance between two areal units  $i$ and $j$ is the path that minimizes the number of ``hops'' or ``jumps'' from cell $i$ to cell $j$. This distance can be calculated by using a symmetric neighboring matrix, where the entry at row $i$ and column $j$ (and vice versa), is $1$ if areas $i$ and $j$ are adjacent, and $0$ otherwise. 
%and that the symmetric matrix $W$ represents the $I \times I$ adjacency relation between cells of the grid. 
%From any two cells $i$ and $j$, we can choose the shortest paths by considering the function defined intuitively by picking the path that minimizes the number of ``hops'' from $i$ to $j$. 
%The logic requires a spatial configuration (in our application it can be %defined through the adjacency matrix) and the distance between two cells $i$ and $j$ is the path that minimizes the number of ``hops'' or ``jumps'' from cell $i$ to cell $j$. 
% On this spatial configuration, we can define various properties that combine different requirements on relevant spatio-temporal quantities.  
While the framework is rather general, we are primarily interested in 
the properties in a predictive context. Thus, we will formulate and consider requirements on the future crowdedness values up to $h$-steps ahead of a given time $t$. Each requirement that we formulate can be checked for every areal unit $i = 1, ..., I$. 
%$\boldsymbol y_{t+1:h}=(\boldsymbol y_{t+1},\ldots,  \boldsymbol y_{t+h})$.
%% Example: given that in the previous 2 hours, the crowdedness was low, the crowdedness shall stay low also in the future 30 min.
% The spatio-temporal signals on which our specification is defined, can be represented by the vector $\mathbf{Y} = [ \boldsymbol{y}_0, \dots, \boldsymbol{y}_h]$, where $\boldsymbol{y}_h$ corresponds to the $h$-step ahead ($I \times T$) predicted crowdedness, with $0$-step representing the current-time observed values.
% :
% \begin{equation*}
%     \boldsymbol{y} : I \times T \rightarrow \mathbb{R}^n
% \end{equation*}
%which is a multi-valued trace that aggregates the levels of crowdedness for any time and location, so that the h-th element $y^h$ of the vector $\boldsymbol{y}$ represents the respective \textit{h-step} ahead log predictive density score prediction, where 0-step denotes the current-time data.
The logic formulae are then specified with the language generated by the following grammar,  which defines rules for building formulae recursively starting from the atomic proposition:
\begin{equation}\label{eq:strel}
    \varphi := \mu~|~\neg~\varphi~|~\varphi_1~\land~\varphi_2
    ~|~\ev{\leq h}~\varphi~|~\glob{\leq h}\varphi
    %~|~ \varphi~\since{[t_1, t_2]}~\varphi
    ~|~\varphi_1~\reach{\leq d}~\varphi_2
    ~|~\escape{\leq d}~\varphi.
    % ~|~    \somewhere{[d,d_2]}{}
    % ~|~ \everywhere{[d_1,d_2]}{}
\end{equation}
The atomic propositions $\mu$ in STREL are defined for a location~$i$ and time~$s$ and
they can describe the indexes (e.g.,~whether location~$i$ contains a hospital 
or whether $s$ corresponds to midnight) or they can be defined as inequalities on the relevant variables
(in our application city crowdedness) e.g.,~$y_{i,s}>c_{i,s}$ or $y_{i,s}<c_{i,s}$ for $c_{i,s}\in\mathbb{R}$.
Note that the quantities entering the atomic propositions are univariate and that the logic cannot, at the
time of writing, express inequalities of the form e.g.,~$y_{i,s} > y_{i,s-1} + c_{i,s}$.
The logical operators then combine different truth values
of atomic propositions $\mathbb{1}(\mu_{i,s}) \in \{0,1\}$ for a sequence of locations and time points, i.e., $\forall (i,s)$ in an index set $\Lambda$.
Boolean operators like $\neg$ and $\land$ denote the classical negation and 
conjunction. We use $\ev{\leq h} \varphi$ and $\glob{\leq h} \varphi$ as the  \emph{eventually} and 
\emph{always} temporal operators, respectively. The former denotes the occurrence of property 
$\varphi$ at least \emph{once} in the future time interval $(t, t + h]$,
while the latter checks the occurrence of property $\varphi$ in \emph{all} future time points
in the interval $(t,t + h]$ (for discretely observed systems, a constant behavior in between time points is
assumed). 
% Note that, in the same fashion, we assume from time~$t$ up to time $t+1$ the process stays 
% constant. This implies that for checking the validity of the temporal properties over the 
% whole interval $(t,t+h]$, $\bm y_t$ (i.e., value at last observed time point) is also considered a relevant quantity. 
% We, however, omit this quantity from the exposition, to not detract the reader from the focus on 
% the predictive aspect of the framework.  
When the context requires also lower bounds to the times 
of interest, we will adopt an interval-based notation, like $\ev{[a,b)}$, to denote the interval $[t+a,t+b)$.
Lastly, spatial operators are represented by the reach $\varphi_1~\reach{\leq d}~\varphi_2$  and escape $\escape{\leq d}~\varphi$
operators for a distance $d\in \mathbb R^+$.
The former represents the \emph{reachability} of an area where $\varphi_2$ holds by only passing through locations
that satisfy $\varphi_1$ where the total distance of the path should be at most $d$. 
The latter operator describes the possibility of
\emph{escaping} from a certain location via a route passing only through locations that satisfy $\varphi$,
with the distance between the starting location of the path and the last being at most $d$. Moreover, other operators such as
disjunction $\land$, implication $\rightarrow$, or further spatial operators can be derived. One example is
the \emph{somewhere} operator $\somewhere{\leq d}~\varphi$ which checks whether there exists a location that
satisfies $\varphi$ reachable via a route with a distance of at most $d$.

%Considering one trajectory of the observed values at time $t$ and draws from the posterior predictive  at time points $t+1,\ldots,t+h$ 
%i.e., $\bmy_{t+1:h}^{(m)}=(\boldsymbol y^o_{t}, \boldsymbol y^{(m)}_{t+1},\ldots,  \boldsymbol y^{(m)}_{t+h})$, 
Once the properties are specified as logic formulae, efficient algorithms tailored to the type of logic employed 
are available to approximate the behavior 
of the stochastic system with respect to the properties. 
One is interested firstly in property satisfaction $S_i(\bm y_{t+1:h}, \varphi)$, i.e., whether the quantities of interest
satisfy property $\varphi$ for area $i$, considering the time period between $t+1$ and $t+h$. It should be noted that the satisfaction of spatial properties in the area $i$ will depend not only on the values of crowdedness in $i$ for the time period of interest, but also on the values in other areas.
STREL provides a Boolean monitoring algorithm for this purpose, which 
returns a yes/no answer while checking for the satisfaction of a given logical formula on a specific 
realization from the system.
Secondly, one also wants to quantify the reliability of a property. This is measured by the robustness function $R_i(\bm y_{t+1:h}, \varphi)$, which is defined as the bound on the perturbation that the quantities of interest can tolerate without 
changing the truth value of a property ~\citep{robustness-fainekos}. The quantitative monitoring algorithm of STREL
computes the value of the robustness function for a given realization. There is a soundness property between the Boolean and the quantitative monitoring such that a positive value corresponds to satisfaction
and a negative value to violation of the property. The robustness function is defined recursively on the operators of the logic \citep[see][]{lmcs_8936}, starting from the atomic proposition. For the atomic proposition, e.g., $\mu_{i,s}=y_{i,s}>c_{i,s}$, the robustness is given by the difference between the quantity of interest $y_{i,s}$ and the threshold value $c_{i,s}$.

We refer the interested reader to~\cite{lmcs_8936} for a complete and formal description of the logic, 
or to~\cite{10.1007/978-3-030-60508-7_2} for a more practice-oriented list of case studies. We devote 
the rest of this section to highlighting the key benefits of adopting the STREL machinery for property verification.
The primary advantage is that STREL is a specification language crafted specifically for keeping a strong 
connection with intuitive notions of spatial and temporal concepts, allowing to express complex requirements in a compact and understandable way. Note that a dedicated scripting language for STREL is
available. It allows expressing the formulae in almost plain English.
Moreover, a key advantage of specifying requirements in terms of STREL operators is that the open-source software Moonlight~\cite{moonlight}
is readily available for automatically verifying that a given set of predictions satisfies the provided specification.
Lastly, the automatic monitoring of STREL specification implemented by Moonlight takes into
account state-of-the-art algorithms for maximizing memory and computational time efficiency 
(with usually better performances than alternatives). While ad-hoc algorithms can be more efficient 
if they are tailored to a given specification, they are often costly to adapt as the
monitored properties evolve. Therefore, the generality offered by the framework and the Moonlight software 
ensures easy adaptability of the property specifications with minimal changes from the modeler.

\subsection{Crowdedness requirements}
\label{sec:properties}

\subsubsection{Informal specification of requirements}

In this section, we propose some informal properties that the crowdedness level in a big city should satisfy to robustly withstand critical events.
Let $c$ represent a crowdedness threshold for all the areas of the city. Note, however, that the framework can accommodate for, e.g., area-specific threshold values. We restrict ourselves to a universal $c$ for the sake of simplicity in the exposition. This threshold would typically be known to the decision-maker and would correspond to the maximum value for which a certain location would still be considered uncrowded.
Moreover, let $h_\varphi$ be a time step in the future to be used in a property or requirement~$\varphi$.

%As in the previous section, we assume that we train our Bayesian model using data up to time $t < T$.

%Given that the measure of crowdedness we are employing is derived from mobile phone data, 
One possible stakeholder of our proposed framework is a telecommunications company, which would like to have a predictive alert system to ensure that their mobile network does not get overcrowded. The following three properties could be of interest to the telecommunications company:
\begin{enumerate}[label=\textbf{P.\arabic*}] 
    \item \emph{Overloads are temporary}: if the level of crowdedness goes above the threshold $c$ in the period following $t$, then it must return below~$c$ latest by time $t + h_{\text{P.}1}$.
    \item \emph{Overloads are local}: if at a certain location  the level of crowdedness at $t + h_{\text{P.}2}$ rises above $c$, this location must be at most at distance $d_{\text{P.}2}$ from
    %to reach 
    %(i.e.,~have a common vertex) to 
    another location  with a level of crowdedness below $c$ at the same time. This is a minimal spatial requirement for a city grid trying to balance excessive loads. 
    \item \emph{The network is fault-tolerant}: for a location, the level of crowdedness in that location or in other locations within a distance of $d_\text{P.3}$ 
    %an adjacent one in the grid  
    should be below $c$ at all times in the interval $(t, t+h_{\text{P.}3}]$, i.e.,~emergency load-balancing must be possible.
\end{enumerate}
In addition to the previous requirements that are related to general aspects of the mobile network,
for the evaluation of the city in terms of safety and quality of life, it is interesting to look at how the city is performing with respect to the reachability of some key points of interest. For example, in an emergency scenario, a traffic monitoring body would be interested in the following requirement (assuming that our crowdedness measure is indeed a proxy for population density in the city):
\begin{enumerate}[label=\textbf{P.\arabic*}] 
\setcounter{enumi}{3}
%\addtolength{\itemindent}{1cm}
    \item \emph{Uncrowded reachability}: A
    %At a future time point $t + h_{\text{P.}4}$  a 
    hospital must be reachable within a distance of $d_{\text{P.}4}$
    from any uncrowded location of the city center, in the time interval $(t,t + h_{\text{P.}4}]$, by only going through uncrowded locations.
\end{enumerate}
%All the requirements will be checked for each location in the grid $i = 1,\ldots, I$.

\subsubsection{Formalizing the requirements}
We show here how to use the STREL logic presented in Section~\ref{sec:strel} 
to specify the requirements introduced above. 
%Description of the properties with the logic.
The previous requirements will be in the following formally expressed as STREL formulae, and the key operators will be described gradually. 
Before looking at the formalization of the requirements, we introduce the atomic property,
\begin{align*}
%\label{atoms}
    %\phi=(y > K),\qquad  \neg \phi=(y < K)
    \phi=(y > c),
\end{align*}
i.e., the crowdedness is above a certain threshold $c$. Conversely, the formula $\neg \phi$ represents the case where the crowdedness level is below or equal the threshold $c$. This formula constitutes the basic building block for formalizing our requirements; in fact, the first requirement is related to temporary overloads, which can be formulated by using temporal operators in the following way:
%uses temporal operators and is specified as:
%
\begin{align}
\label{p1}\tag{P.1} \varphi_{P.1} = \phi \rightarrow \ev{\leq h_{P.1}}\neg \phi.
%, \qquad \phi=(y > K),\qquad  \neg \phi=(y < K)
\end{align}
%
%~$\ev{\leq h_{P.1}}{}\neg \phi$  represents the temporal ``eventually'' operator that requires the property $\neg \phi$ to hold for at least one-time unit within the interval 
%$[0, h_{P.1}]$ where the convention is that $0$ is the beginning of the testing period; lastly, ``$\rightarrow$'' represents the implication operator.
% meaning that the right side must hold when the left side does. 
% Therefore, the meaning of the whole formula \ref{p1} literally would be: if the crowdedness is above K then it will decrease below K within the next 3 time units.
%%
% Applications aimed at identifying the locations least likely to meet the requirement of \ref{p1} could consider a stricter version of \ref{p1} by imposing that the property should be true for any $t$ (i.e., regardless of the training sample considered). This can be specified by the property $\varphi_{P.1'}$ using the  ``globally''  operator: 
%
% \begin{align}
% \label{p1'}\tag{P.1'} % \glob{}\varphi_{P.1}.
%%= \phi \rightarrow \ev{\leq h_{P.1}}\neg \phi
%, \qquad \phi=(y > K),\qquad  \neg \phi=(y < K)
% \end{align}
%
% When we omit the subscript to temporal operators, we mean they must hold for all training samples $t$. So, in our case, it requires the property $\phi$ to hold \emph{for all} the time instants.
%

The second property, related to overloads being local, can be formalized as a spatio-temporal property: 
\begin{align}
 \label{p2}\tag{P.2} \varphi_{\text{P.}2} = \ev{= h_{\text{P.}2}} (\phi \rightarrow  %\ev{=h_{\text{P.}2}} (
 \somewhere{\leq d_{\text{P.}2}}{} \neg \phi).
\end{align}
%The  ``somewhere'' operator $\somewhere{\leq d_{\text{P.}2}}{} \neg \phi$ denotes the fact that there must be a location at the distance of $d_{\text{P.}2}$ grid square
%(when $d_{\text{P.}2}=1$), where $\neg \phi$ holds. 
Note that the temporal operator $\ev{=h_{\text{P.}2}}$ here denotes that the requirement shall hold at the time which lies $h_{\text{P.}2}$-steps ahead in the future.

Thirdly, the fault-tolerance of the network is a spatio-temporal requirement, which can be formulated as
\begin{align}
\label{p3}\tag{P.3}  \varphi_{\text{P.}3} =  \glob{\leq h_{\text{P.}3}} \left(\somewhere{\leq d_{\text{P.}3}}{}  \neg \phi\right),
\end{align}
where $\glob{\leq h_{\text{P.}3}}$ requires the ``somewhere'' property to hold globally for the whole interval $(t, t+h_{\text{P.}3}]$. 
% The overall property \ref{p3} can therefore be read in literal English as if there is a location where the level of crowdedness stays consistently above the threshold K for three time-units, then there must be, during that time, a nearby location having a level of crowdedness below the threshold.

Lastly, for the requirement related to the reachability of hospitals, let us first introduce a new atomic proposition:
%in the city could be written as: 
%%
\begin{align*}
\phi_2 = isHospital.
\end{align*}
Here, $\phi_2$ is a peculiar proposition representing hospital locations, meaning that it is satisfied only when the current location comprises a hospital. 
A first, direct translation of the P.4 could be the following:
\begin{align*}
\ev{\leq h_{\text{P.}4}} (\neg \phi \reach{\leq d_{\text{P.}4}}{} \phi_2).
\end{align*}
While the previous requirement formalizes P.4 \emph{literally}, it likely gives an unrealistic interpretation of the requirement. In fact,  $\neg \phi \reach{\leq d_{\text{P.}4}}{} \phi_2$ means that a hospital can be reached by only traversing uncrowded areas, however, it does not consider the traveling time to reach the location, meaning that the property would be satisfied \emph{whenever} there is an uncrowded path to the hospital (although the actual traveling time might be significantly higher).
A more realistic (although a bit more involved) version is presented below:
%%
% \begin{align}
% \label{p4old}\tag{P.4}  \varphi_{\text{P.}4} = \ev{\leq h_{\text{P.}4}} (\neg \phi \reach{\leq d_{\text{P.}4}}{} \phi_2),\qquad \phi_2 = isHospital
% \end{align}
%
\begin{align}
\label{p4}\tag{P.4}  \varphi_{\text{P.}4} = \psi_{0,d_{\text{P.}4}}, 
%,\qquad \phi_2 = isHospital
\end{align}
\begin{align*}
 \psi_{i,n} = 
 \begin{cases}
 \phi_2 & n = 0,\\
 \phi_2 \vee   \left(\left(\glob{[i,i+1]} \lnot \phi \right) \wedge \left( \somewhere{\leq 1}{}\psi_{i+1,n-1} \right)\right) & \mathit{otherwise}.
 \end{cases}
\end{align*}
%
% \begin{align*}
%  \varphi_{\beta,j} = 
% \begin{cases}
%   \phi_2 & j = 0\\
%   \psi_{i+1}, & \mathrm{otherwise}
% \end{cases}
% \end{align*}
%
% \begin{align}
% \label{p4}\tag{P.4}  \varphi_{\text{P.}4} = \phi_2 \vee \left(\bigvee\limits_{i=0 \dots d_{\text{P.}4}-1} \psi_{i,i}\right)
% %,\qquad \phi_2 = isHospital
% \end{align}
% %
% \begin{align*}
%  \psi_{n,j} = \left(\glob{=n-j} \lnot \phi \right) \wedge \left( \somewhere{\leq 1}{}\varphi_{\beta,n,j} \right)
% \end{align*}
% %
% \begin{align*}
%  \varphi_{\beta,n,j} = 
% \begin{cases}
%   \phi_2 & j = 0\\
%   \psi_{n,j-1}, & \mathrm{otherwise}
% \end{cases}
% \end{align*}
% where $\glob{=0} \phi$ denotes the evaluation of $\phi$ at the beginning of the testing period.
Property $\varphi_{\text{P.}4}$, with a slight abuse of notation, encodes our requirement and shows the flexibility of the logic approach. 
%Although a little complex to grasp, $\varphi_{\text{P.}4}$ shows the flexibility of the logic approach.
The requirement states that one needs to move at most  $d_\text{P.4}$ cells in the time interval $(t, t+h_\text{P.4}]$ time units, but it does not explicitly specify how fast one can move through the different cells of the city grid. To give a realistic interpretation of the specification, we assumed that in one time unit (i.e., $10$ minutes), one can only travel from one cell to the next. This interpretation translates into $\psi_{  j, n}$, which imposes that the current location is not crowded for the next ten minutes, and iteratively enforces this $n$ times by the recursive check of $\psi_{ j, n}$ until the maximum distance is reached (in terms of ``hops'' on the grid), in which case it looks for a hospital in the neighborhood. This way, analyzing the satisfaction or robustness of $\varphi_{\text{P.}4}$, will not only provide insights about the spatial reachability of a hospital, but it will also take into account the traversal time needed to reach it.

\section{Predictive model checking and model comparison using formal methods}
\label{sec:bayes}
\subsection{Incorporating uncertainty through the Bayesian approach}
\label{sec:uncertainty}
Our application aims to study the behavior of city crowdedness observed at regular time intervals on a fixed grid area. Again, let $y_{i,j}$ denote the crowdedness measure in area~$i$ at time~$j$ for $i=1,\ldots I$ areal units on a city grid and $j=1,\ldots T$ time points.  As a framework for predictive inference, we assume the observations up to time $t < T$ are used as a training sample and the evaluation is performed on the observations at the remaining $t+1, \ldots, T$ discrete time points. By using STREL, we are interested in investigating two functions or statistics of future crowdedness, namely
the property satisfaction and the property robustness.  
To extend the concept of satisfaction and robustness over the whole stochastic system, we introduce two key concepts:
i) the Bayesian predictive probability of satisfaction and ii) the expected value of the  Bayesian predictive robustness.

As mentioned in the introduction, traditional SMC considers the parameters of the model fixed, generates a large number of trajectories from the model and then considers how many times the trajectories satisfy a given property. 
In the Bayesian framework, the parameters are not fixed but are given a prior distribution based on expert knowledge or prior observations.  After observing the data, one then obtains a posterior probability distribution for the model parameters. 
We now show how we use the Bayesian approach to account for uncertainty on the parameters when calculating the probability of satisfaction and the robustness of a property. 
We start by introducing the Bayesian predictive density. 
The $h$-step ahead Bayesian predictive density is given by
\begin{equation}\label{eq:preddens}
    p(\boldsymbol y_{t + h}| \boldsymbol y^o_{1:t})=\int_{\mathcal{K}}p(\boldsymbol y_{t + h}| \boldsymbol y^o_{1:t}, \boldsymbol \kappa) p(\boldsymbol\kappa|\boldsymbol y^o_{1:t})d\boldsymbol\kappa,
\end{equation}
where $\boldsymbol y^o_{1:t}$ denotes the observed values up to time $t$ of random variables $\boldsymbol y_{1:t}=(\boldsymbol y_1, \ldots, \boldsymbol y_t)$ (each $\bm y_{.}$ being $I$~dimensional), $\boldsymbol\kappa$ contains all parameters and latent quantities to be estimated in the model, $p(\boldsymbol\kappa|\boldsymbol y^o_{1:t})$ denotes their posterior distribution and $\mathcal{K}$ contains the corresponding integration space. It can be seen that the predictive density in Equation~\eqref{eq:preddens} is given by the integral of the likelihood function, where the values of the unobservables $\boldsymbol\kappa$ are weighted according to their posterior distribution. This means that this predictive density integrates uncertainty about the vector of unobservables and the intrinsic uncertainty about the future value $\boldsymbol y_{t+h}$ given the history $\boldsymbol y^o_{1:t}$.
The posterior distribution $p(\boldsymbol\kappa|\boldsymbol y^o_{1:t})$ can, in our proposed models, be accessed by generating $M$ draws  
$\boldsymbol\kappa_{1:t}^{(m)}$ from the posterior up to time $t$ using MCMC. The predictive distribution in Equation~\eqref{eq:preddens} can then be accessed by simulating $\boldsymbol y^{(m)}_{t+h}$ from each of the distributions represented by the density $p(\boldsymbol y_{t + h}| \boldsymbol y^o_{1:t},\boldsymbol\kappa_{1:t}^{(m)})$ for $m = 1,\ldots,M$.

%Aside from classical posterior predictive assessment, one can also employ existing techniques from the computer science field of formal verification methods to formulate and automatically verify relevant properties for the system being modeled. 
% In the following we show how the formal verification methods introduced above can be utilized in a natural 
% way in a context where the stochastic system is defined following the Bayesian paradigm.
% and how such methods 
% complement posterior predictive inference and model checking.
%The framework can help statisticians and decision-makers to automatically check whether the system will satisfy or not these properties at a future time point \emph{a posteriori}.

Starting from the predictive distribution, we can now define the 
Bayesian $h$-step-ahead predictive probability of satisfaction for property $\varphi$ at location~$i$ and time $t$ by
\begin{align}\label{eq:int_mcmc}
 \mathbb{E}&[S_i(\bm y_{t+1:h}, \varphi)|\bm y^o_{1:t}]= \nonumber\\
& \int_{\bm y_{t+1} \in \mathcal{Y}_1}\ldots \int_{\bm y_{t+h} \in \mathcal{Y}_h} S_i(\bm y_{t+1:h}, \varphi)p({\bm y_{t+1:h}}| \bm y^o_{1:t})\text{d}{\bm y_{t+1}}\ldots\text{d}{\bm y_{t+h}}.
 \end{align}
%In the above expression $p({\bm y_{t+1:h}}| \bm y^o_{1:t})$ is referred to as the predictive distribution for the future observations $\bm y_{t+1:h}$ given the observations up to time $t$, that is $\bm  y^o_{1:t}$. 
%Let  $\bm \theta$ be the vector of parameters of our model, which can take values
%$\bm \theta \in \Theta$ and $p(\bm \theta)$ its prior %probability distribution, we can express the predictive probability as
%\begin{align*}\label{eq:int_pred}
% p({\bm y_{t+1:h}}| \bm y^o_{1:t}) = 
% \int_{\bm \theta \in \Theta}p({\bm y_{t+1:h}}| \bm y^o_{1:t}, %\bm \theta) p(\bm \theta| \bm y^o_{1:t}) \text{d}{\bm \theta},
% \end{align*}
% where  $p(\bm \theta| \bm y^o_{1:t})$  is the posterior predictive probability of the vector parameter $\bm \theta$ and according to the Bayes' theorem is proportional to $p(\bm \theta) p(\bm y^o_{1:t}| \bm  \theta)$.
% Now that we have shown that the Bayesian framework allows incorporating the uncertainty around the model parameters in the formulas for the probability of satisfaction, we need to calculate the probability of satisfaction. 
% This quantity is computed conditional on the observed data and takes into account the two uncertainty components discussed above.
We can avoid the calculation of the multidimensional integral in Equation \ref{eq:int_mcmc} and approximate the probability by using the draws from the Bayesian predictive density,
\begin{equation}
\label{eq:satisf_sum_mcmc}
\mathbb{E}[S_i(\bm y_{t+1:h}, \varphi)|\bm y^o_{1:t}]\approx \frac{1}{M}\sum_{m=1}^MS_i(\bm y_{t+1:h}^{(m)}, \varphi),
\end{equation}
where for one draw $m$ the $S_i(\bm y_{t+1:h}^{(m)}. \varphi)$ takes either value zero or value one.

The Bayesian predictive robustness is a function of the relevant predictions
for the property $\varphi$ together with the property parameters. For each location $i$ and time $t$ we can obtain a distribution of this measure by integrating over the posterior distribution of the unknowns. One can then look at summary statistics of this distribution, such as the expected value at location $i$ and time $t$, which is given by
\begin{align}\label{eq:robust_mcmc}
 \mathbb{E}&[R_i(\bm y_{t+1:h}, \varphi)|\bm y^o_{1:t}]= \nonumber\\
 &\int_{\bm y_{t+1} \in \mathcal{Y}_1}\ldots \int_{\bm y_{t+h} \in \mathcal{Y}_h} R_i(\bm y_{t+1:h}, \varphi)p({\bm y_{t+1},\ldots, \bm y_{t+h}}| \bm y^o_{1:t})\text{d}{\bm y_{t+1}}\ldots\text{d}{\bm y_{t+h}},
 \end{align}
 and can be approximated by
\begin{equation}
\label{eq:robust_sum_mcmc}
\mathbb{E}[R_i(\bm y_{t+1:h}, \varphi)|\bm y^o_{1:t}]\approx \frac{1}{M}\sum_{m=1}^MR_i(\bm y_{t+1:h}^{(m)}, \varphi).
\end{equation}
Note, however, that also other summary statistics of the distribution could be employed for decision-making. For example, if a more conservative approach is desired, one can consider higher quantiles (e.g.,~90\%) of the posterior distribution of the robustness.

The monitoring algorithms of STREL are employed here to efficiently calculate \eqref{eq:satisf_sum_mcmc} and \eqref{eq:robust_sum_mcmc} 
from the $m=1,\ldots, M$ posterior predictive draws. We note that both $S(\bm y_{t+1:h}, \varphi)$ and $R(\bm y_{t+1:h}, \varphi)$ 
can be seen as summary statistics in the sense of the classical posterior predictive model check \citep[PPC,][]{gelman1996posterior}. In our case,
unlike in classical PPC, the summary statistics do not depend on the model parameters, but rather on the property parameters which are kept fixed
throughout the analysis. The satisfaction is a binary statistic, while the robustness is continuous. Methods such as Bayesian p-values \citep{gelman1996posterior} can be employed on these statistics for the purpose of model evaluation.

\subsection{Property-driven model comparison}
\label{sec:property_comparison}

%In this section, we introduce the framework for performing model comparisons using formal methods.
%We begin by presenting the Bayesian predictive distribution and the log predictive density scores as classical ex-post predictive evaluation measures. We then introduce formal verification methods in general and spatio-temporal reach and escape logic \citep[STREL,][]{lmcs_8936} 
%as the language used for specifying properties in particular.
%We conclude this section by introducing two Bayesian posterior predictive quantities derived from the formal verification method, namely the posterior predictive satisfaction and robustness.

% Not only does the Bayesian predictive distribution in~\eqref{eq:preddens} have the advantage of combining the two sources of uncertainty in a coherent framework but the use of simulation methods to produce $\boldsymbol\kappa_{1:t}^{(m)}$  and then $\boldsymbol y^{(m)}_{t+h}$ makes this predictive distribution applicable in real-time and hence feasible in a smart city context. \todo[author=LV]{incorporate with following section}

The Bayesian predictive distribution in~Equation~\eqref{eq:preddens} can also be employed for the purpose of model comparison by using the $h$-steps ahead log predictive density scores \citep[cf.][]{geweke2010comparing, kastner2016dealing}.
If we evaluate~\eqref{eq:preddens} at the observed value $\boldsymbol y^o_{t+h}$, the $h$-step ahead LPDS is the real number:
$$
\text{LPDS}_{t+h} = \log \int_{\mathcal{K}}p(\boldsymbol y^o_{t + h}| \boldsymbol y^o_{1:t}, \boldsymbol \kappa) p(\boldsymbol\kappa|\boldsymbol y^o_{1:t})d\boldsymbol\kappa\approx \log\left(\frac{1}{M}\sum_{m=1}^{M}p(\boldsymbol y^o_{t + h}| \boldsymbol y^o_{1:t}, \boldsymbol\kappa_{1:t}^{(m)})\right).
$$
% Using the LPDS as an evaluation measure instead of classical methods such as the root mean squared error (RMSE) has the advantage of accounting for the distribution in the prediction while RMSEs neglect the uncertainty surrounding the point forecasts.
%%
The LPDS evaluates a predictive model based only on the
density value at the realizing outcome. However, it is not the only metric that can be employed for evaluating the predictive performance of probabilistic forecasts. 
Other scores can be employed depending on which forecast feature is desirable for the application at hand. For example, another commonly
employed score which also rewards predictive distributions that place mass close to the realizing outcome is the continuous-ranked probability score \citep[CRPS;][]{Matheson1976}.  
More generally, these scores are sample estimates based on the observed data of scoring rules employed to measure prediction accuracy 
\citep[][]{gneiting2007strictly}. For a review of the estimation of scoring rules based on MCMC output, see \cite{Krueger2021}.

We propose that, in addition to the above methods well-established in the Bayesian Machine Learning community,  model
assessment and comparison should be enhanced with considerations specifically tailored for the application at hand, and that they should take into account how the
predictions from the Machine Learning model are used for decision making. This is why in the following we propose several measures that rely on the posterior predictive satisfaction and robustness of the properties introduced in Section~\ref{sec:uncertainty}.

In particular, we compare the posterior predictive satisfaction and robustness measures estimated on the $M$ trajectories with the ex-post evaluation of the satisfaction and robustness of the properties on the observed data after time~$t$ $\bm y^o_{t+1:h}$.
We compute for the following measures: 
\begin{itemize}
    \item Mean accuracy between the observed and estimated satisfaction,
$$
\bar{\text{Acc}}_t^{\text{Satisf}} =  \frac{1}{M}\sum_{m=1}^{M}\left(\frac{1}{I}\sum_{i=1}^{I}1\{S_i(\bm y_{t+1:h}^{(m)}, \varphi)=1\} 1\{S_i(\bm y^o_{t+1:h}, \varphi) = 1\}\right).
$$  
    \item Mean F1 score between the observed and estimated satisfaction,
$$
\bar{\text{F1}}_t^{\text{Satisf}} =  \frac{1}{M}\sum_{m=1}^{M}2\frac{\text{recall}^{(m)} \times \text{precision}^{(m)}}{\text{recall}^{(m)} + \text{precision}^{(m)}},
$$  
where
$$\text{precision}^{(m)}=\frac{\sum_{i=1}^{I}1\{S_i(\bm y_{t+1:h}^{(m)}, \varphi)=1\} 1\{S_i(\bm y^o_{t+1:h}, \varphi)=1\}}{\sum_{i=1}^{I}1\{S_i(\bm y_{t+1:h}^{(m)}, \varphi)=1\}},$$
$$
\text{recall}^{(m)}=\frac{\sum_{i=1}^{I}1\{S_i(\bm y_{t+1:h}^{(m)}, \varphi)=1\} 1\{S_i(\bm y^o_{t+1:h}, \varphi)=1\}}{\sum_{i=1}^{I}1\{S_i(\bm y^o_{t+1:h}, \varphi)=1\}}.
$$
  \item Root mean squared error (RMSE) between the observed and the estimated robustness,
    $$
\text{RMSE}^{\text{Rob}}_t =  \sqrt{ \frac{1}{M}\sum_{m=1}^{M}\left\{\frac{1}{I}\sum_{i=1}^{I}(R_i(\bm y_{t+1:h}^{(m)}, \varphi) - R_i(\bm y^o_{t+1:h}, \varphi))^2\right\}}.
$$
\end{itemize}

%
% For example, something like: 
%
% $$
% \sum_{i=1}^I  \frac{s(\bm y^h_{t}, \varphi, i) - \mathbb{E}[s(\bm y^h_{t}, \varphi, i)|\bm y^o_{1:t}]}{\mathbb{V}[s(\bm y^h_{t}, \varphi, i)|\bm y^o_{1:t}]}, \qquad \sum_{i=1}^I  \frac{R(\bm y^h_{t}, \varphi, i) - \mathbb{E}[R(\bm y^h_{t}, \varphi, i)|\bm y^o_{1:t}]}{\mathbb{V}[R(\bm y^h_{t}, \varphi, i)|\bm y^o_{1:t}]}
% $$
%

%%%%%%%%%%%%%%%%%%%%%%%%%%%%%%%%%%%%%%%%%%%%%%
\section{Empirical illustration}\label{sec:results}
\subsection{Data}\label{sec:data}
To illustrate the proposed framework, we employ a data set 
containing telecommunications activity data derived from
call detail records (CDRs) for the center of Milan, Italy
over the period of one week in November 2013. 
The CDR data is a valuable proxy for population distribution and 
people's mobility habits \citep{peters2017dynamic}, given
the almost universal use of mobile phones, and 
has a high potential in researching the patterns in mobility at a high frequency in time and over a large spatial network.
% The almost universal use of mobile phones has generated vast amounts of data mainly in the form of mobile phone location data through the so-called call detail records (CDRs), which are a valuable proxy for population distribution and people's mobility habits \citep{peters2017dynamic}.
Mobile communication service providers generate a CDR whenever a device state changes either because of the user's actions (phoning, texting,  browsing on the internet) or because of technical reasons (e.g., switching to a cell with a stronger signal in the cellular network).
% The CDR will then record the time and the cell that handled the interaction, data which in turn can be used to obtain an approximate location of the user. This type of information has a high potential in researching the patterns in mobility at a high frequency in time and over a large spatial network.

The data set employed in this paper is a subset of the 
``Telecom Italia Big Data Challenge'' open source database, %within the scope of the ``Telecom Italia Big Data Challenge'',
which contains various geo-referenced, aggregated and anonymized datasets for the city of Milan and the Province of Trentino 
% with the ``main'' data set containing telecommunications activity data derived from CDRs  
\citep[for a detailed description, see][]{barlacchi2015multi}. 
% In general, the acquisition of mobile phone data from telecommunications providers is a rather difficult process primarily due to security and
% privacy concerns and very few open datasets are available for research. 
% Tne exception is provided by Telecom Italia's open source database within the scope of the ``Telecom Italia Big Data Challenge'', which contains various geo-referenced, aggregated and anonymized datasets for the city of Milan and the Province of Trentino, with the ``main'' data set containing telecommunications activity data derived from CDRs  \citep[for a detailed description see][]{barlacchi2015multi}. 
% The datasets are released under the Open Database License (ODbL) and are publicly available in the Harvard Dataverse. 
The telecommunications activity data covers the period November 01, 2013, to December 16, 2013, and the CDR data provided is aggregated in both space and time. In the case of Milan, the city area is composed of a grid overlay of 1000 squares with the size of approximately 235m$\times$235m with the CDRs being aggregated inside each square. Additionally, a temporal aggregation is performed in time slots of ten minutes. Information on the type of activity that generated the CDR is also provided in the database:
\begin{itemize}
    \item SMS-in activity: activity proportional to the amount of received short message services (SMSs) inside a given grid square during a given time interval.  A CDR is generated each time a user receives an SMS.
    \item SMS-out activity: activity proportional to the amount of sent SMSs inside a given grid square during a given time interval.
        A CDR is generated each time a user sends an SMS.
    \item Call-in activity: activity proportional to the number of received calls inside a given grid square during a given time interval.
        A CDR is generated each time a user receives a call.
    \item Call-out activity: activity proportional to the number of issued calls inside a given grid square during a given time interval.
    A CDR is generated each time a user issues a call.
    \item Internet traffic activity: number of CDRs generated inside a given grid square during a given time interval.
     A CDR is generated each time a user starts an Internet connection or ends an Internet connection. During the same connection, a CDR is generated if the connection lasts for more than 15 min or the user transfers more than 5 MB.
\end{itemize}
The data is further anonymized by dividing the true number of records in each category by a constant known to Telecom Italia, which hides the true number of calls, SMS and internet connections.
%%%%%%%%%%%%%%%%%%%%%%%%%%%%
\paragraph{Sample description}
%%%%%%%%%%%%%%%%%%%%%%%%%%%%

We consider a subset of the Milan telecommunications activity dataset covering the period November 04, 2013 (Monday) to November 11, 2013 (Monday) with $T=6 \times 24 \times 8=1152$. 
We further restrict our analysis to the central $21\times 21$ grid (i.e., $I=441$), where the center-most cell is the one containing the location of the Milan Duomo. This grid corresponds to an area of around 25$\text{km}^2$. Moreover, we consider the sum of all the mobile phone activity measures (i.e., SMS-in, SMS-out, call-in, call-out and internet) as our measure of crowdedness. We use this aggregated measure because i) the phoning (call-in, call-out) and texting by SMS measures are rather sparse during the night, as people rarely call or text after midnight, ii) the modern use of cell phones relies much more on browsing the internet or on messaging apps which gained popularity around 2010. As such, considering the internet CDRs in addition to the other four can paint a more realistic picture of the crowdedness of a certain area. 

\begin{figure}[tp]
    \centering
    \includegraphics[width=0.8\columnwidth]{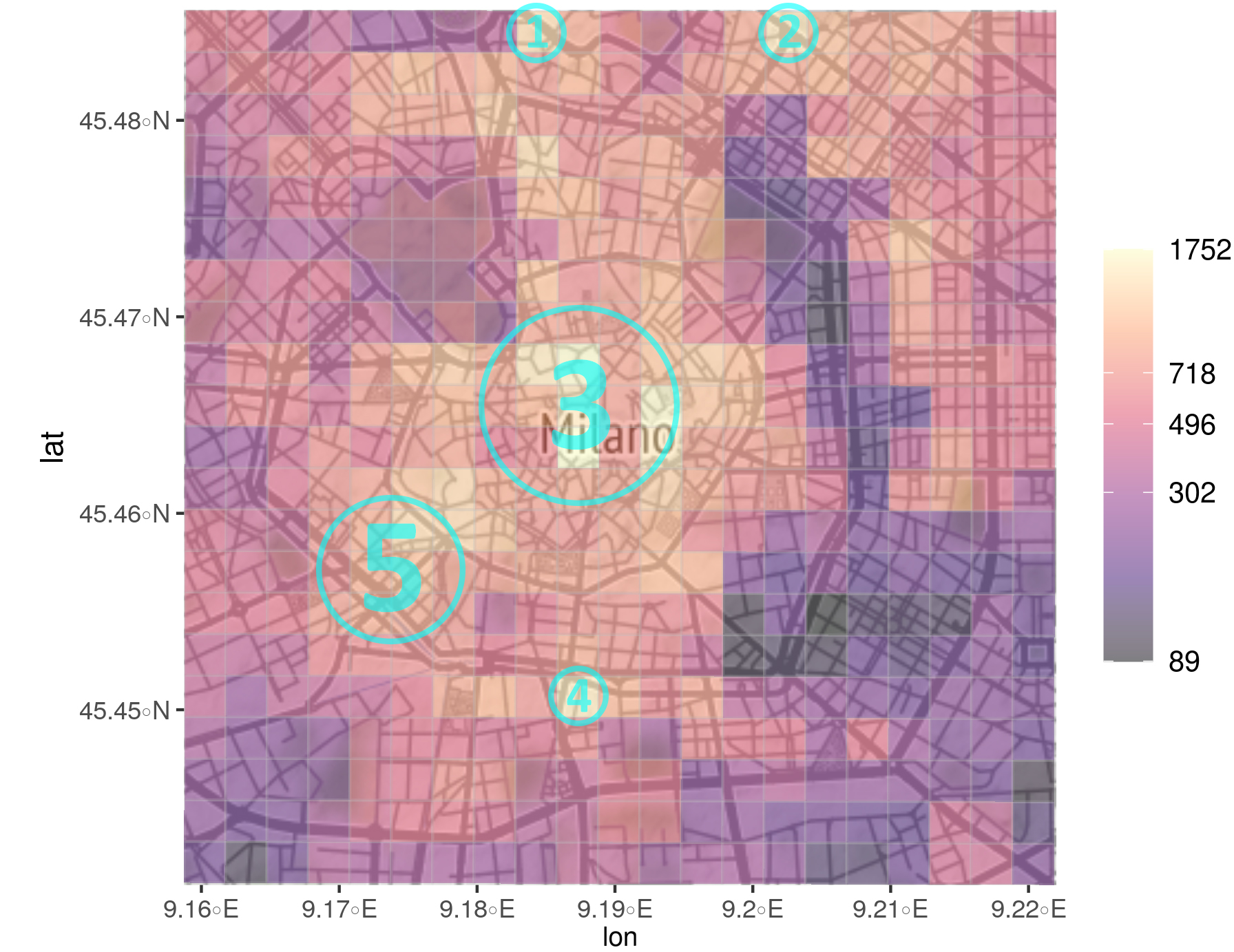}
    \caption{Average crowdedness measure over the period November 04, 2013, to November 11, 2013. The $x$-axis contains the longitude and the $y$-axis contains the latitude degrees. The marked areas represent: \circled{1}~\emph{Garibaldi Station}, \circled{2}~\emph{Central Station}, \circled{3}~\emph{Duomo}, \circled{4}~\emph{Bocconi}, \circled{5}~\emph{Navigli}.}
    \label{fig:agg_grid}
\end{figure}

Figure~\ref{fig:agg_grid} contains the crowdedness measure averaged over the 10-minute time intervals for the whole analyzed period over the $21\times 21$ grid. Areas with high levels of crowdedness are apparent in the central grid squares in the area surrounding the Milan \emph{Duomo} \circled{3} and in the upper center, where the two main stations are, namely \circled{1}~\emph{Garibaldi Station} and \circled{2}~\emph{Central Station}. On the other hand, lower activity grid squares such as the ones overlaid on a highly trafficked avenue on the right-hand side and right bottom corner of the map can be identified.

\begin{figure}[tp]
    \centering
    \includegraphics[width=0.9\columnwidth]{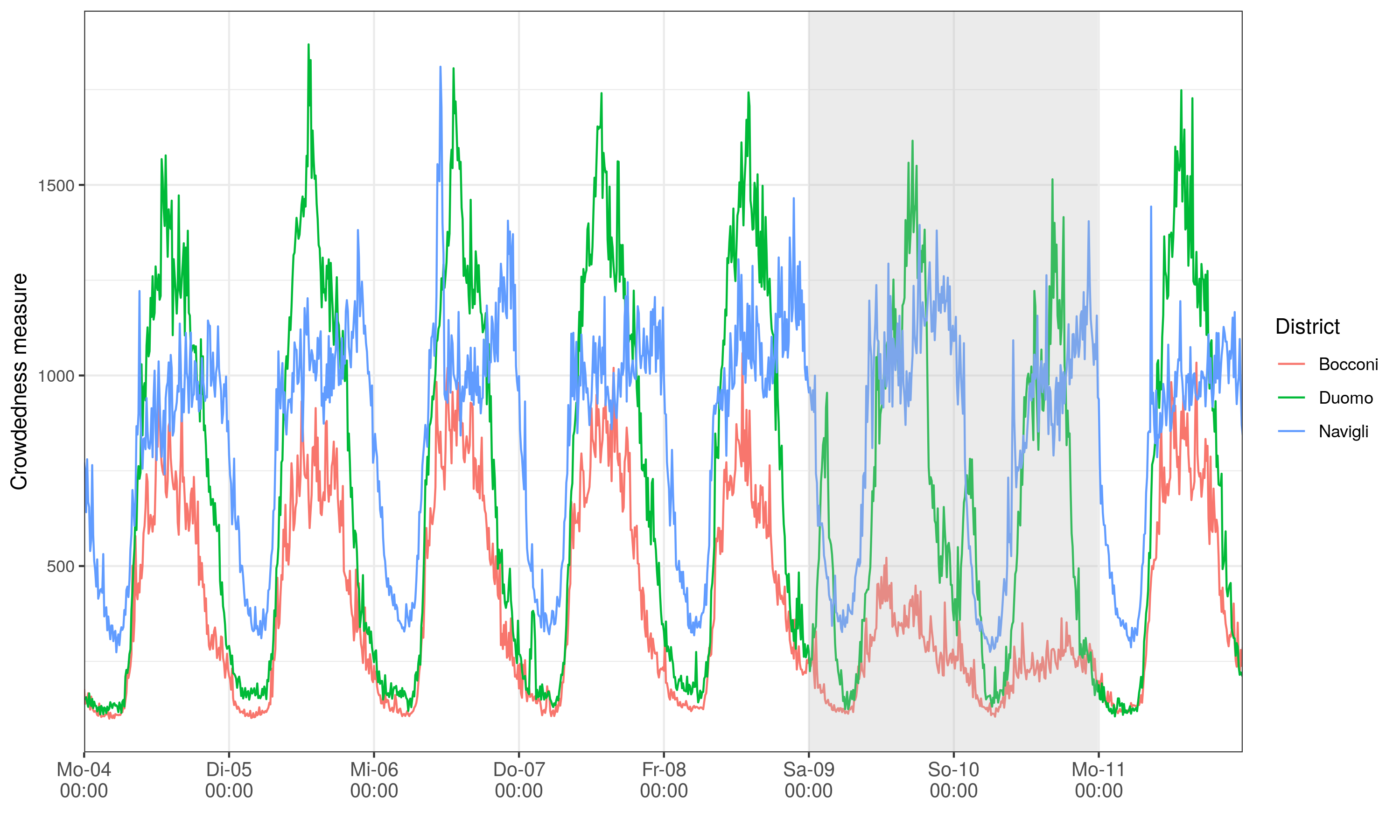}
    \caption{Time series of the crowdedness measure for the areal units containing the districts Duomo, Navigli and Bocconi over the period from November 04, 2013, to November 11, 2013. Areas marked in gray represent weekends.}
    \label{fig:ts_duomo_navigli_bocconi}
\end{figure}

In order to illustrate the temporal behavior of the crowdedness measure, we present in Figure~\ref{fig:ts_duomo_navigli_bocconi} the time-series of the grid units containing the three representative districts of \circled{3}~\emph{Duomo}, \circled{5}~\emph{Navigli} and \circled{4}~\emph{Bocconi}. We observe a larger high activity in the area of \circled{3}~\emph{Duomo} compared to the other two districts, which peaks around midday during the working days and in the early afternoon on the weekends. \circled{5}~\emph{Navigli} on the other hand, which is a district famous for its different types of cafés, restaurants, bars and design shops, exhibits a more uniform behavior among the working and the weekend days, with activity peaking in the evening hours. 
The grid square containing the \circled{4}~\emph{Bocconi} university exhibits a clear pattern during working hours and reduced activity levels on the weekends, especially on Sundays.

\begin{figure}[tp]
    \centering
    \includegraphics[width=0.9\columnwidth, trim=0 20 0 20,clip]{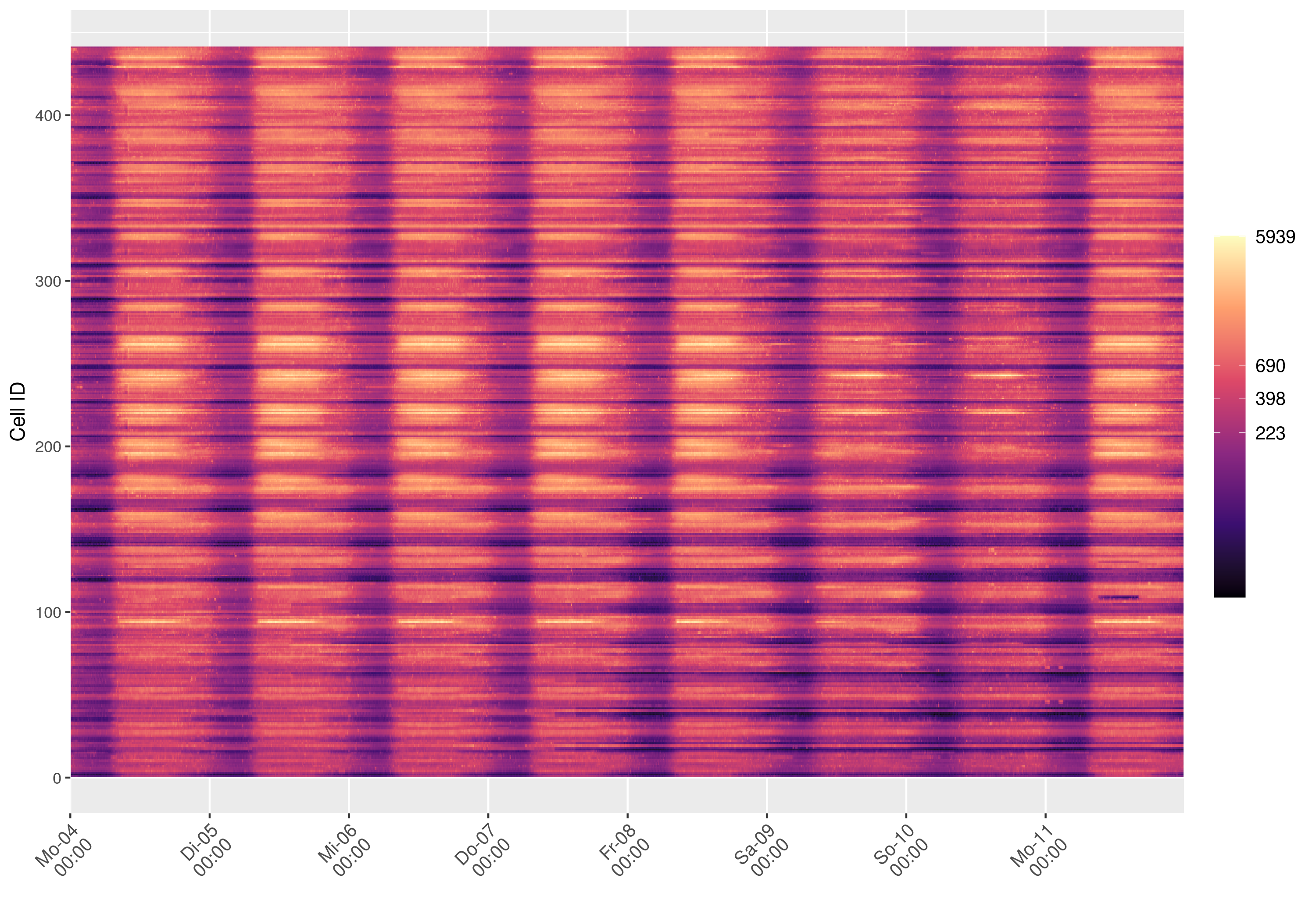}
    \caption{Raster plot exhibiting the time series of the crowdedness measure for each of the 441 grid squares for the period November 04, 2013, to November 11, 2013.}
    \label{fig:raster_st}
\end{figure}

The seasonality in the data can be identified also in Figure~\ref{fig:raster_st}, which contains a visualization of the whole dataset through a raster plot. Aside from the strong daily seasonality that is present in all locations, one can observe different temporal patterns among the areas. Most areas share the characteristic of relatively lower activity on the weekends, while the activity on the working days differs among groups of locations e.g., the locations from the central part of the grid (cell IDs 150--280) exhibit a higher difference between the daily crowdedness during working days vs. weekends, while locations with cell IDs around 400 (top cells in the map) exhibit rather similar activity during the workdays and weekends.

\subsection{Model comparison} \label{sec:predictive-performance}

\begin{figure}[tp]
    \centering
    \includegraphics[width=0.8\columnwidth,trim=0 20 0 5,clip]{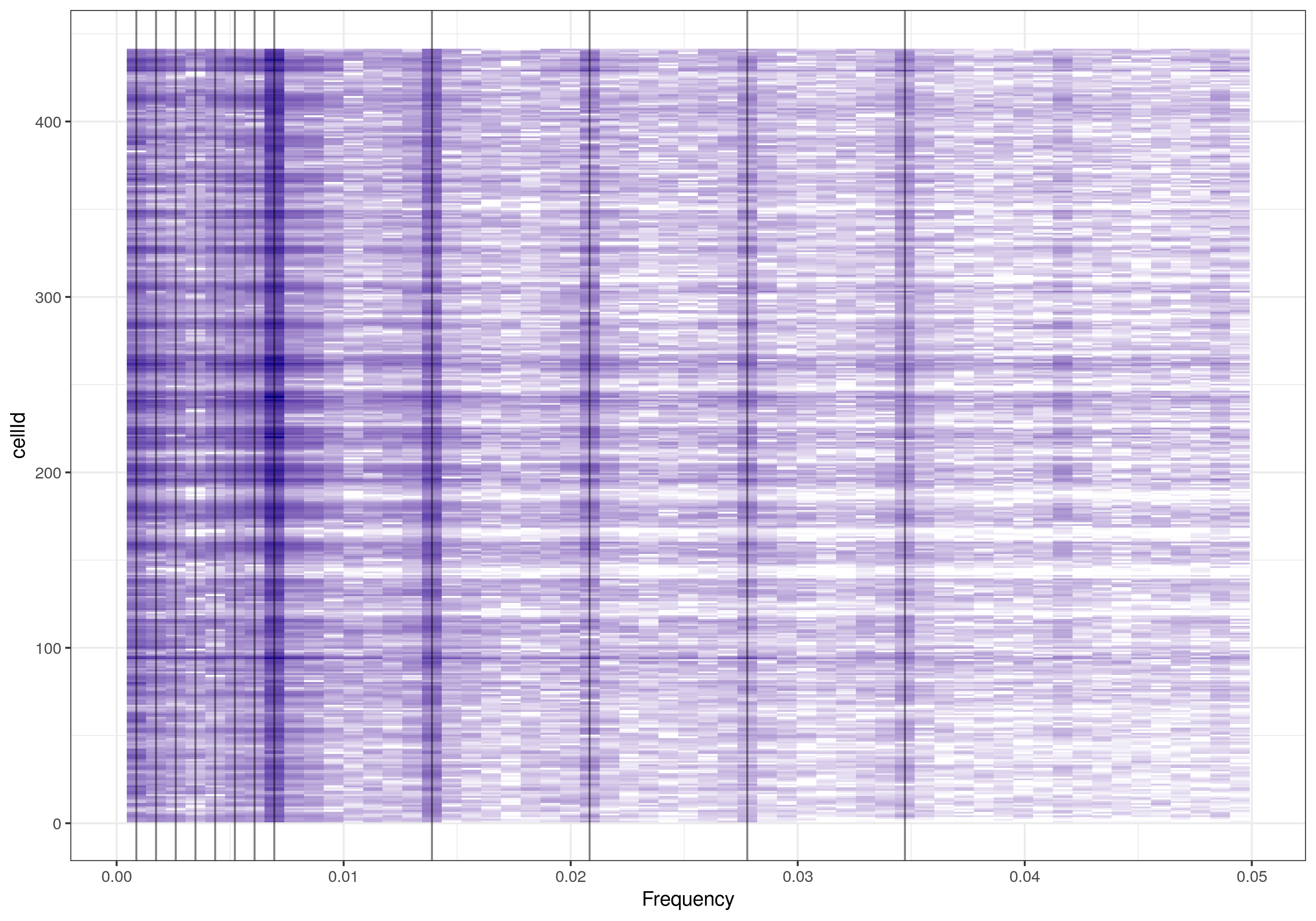}
    \caption{Raster plot exhibiting the spectral density estimates for each of the 441 grid squares for the period from November 04, 2013, to November 11, 2013. The $x$-axis only exhibits frequencies up to 0.05. The black vertical lines represent the frequencies that were chosen to be included in the harmonic regression after visual inspection of this graph. 
    The chosen frequencies correspond to the intra-weekly and intra-daily seasonality.}
    \label{fig:ts_spectral}
\end{figure}

We employ in our analysis several Bayesian models to model the behavior of city crowdedness observed at regular time intervals on a fixed grid area. Let $y_{i,t}$ denote the crowdedness measure in area~$i$ at time~$t$ for $i=1,\ldots I$ areas and $t=1,\ldots T$ time points. 
As could be seen in descriptive figures, the measure exhibits 
seasonal behavior on both a daily and weekly level and this behavior is 
heterogeneous across the different areas. 
%Crowdedness will peak in most areas at noon or in the evening and will drop significantly during the night. Moreover, certain areas will exhibit high activity during the weekdays while others will get crowded during the weekend. 
Similar to \cite{modelac}, we account for such characteristics in the model formulation by employing the following dynamic harmonic regression:
$$
\log(y_{i,t})= \beta_0 + 
\boldsymbol h_t^\top  \boldsymbol\beta_{i} + \eta_{i,t},
$$
where $\boldsymbol\beta_{i}$ is a vector of  area-specific regression coefficients, $\eta_{i,t}$ is an error term and $\boldsymbol h_t$
is a vector of dimension $2K$ of harmonic regressors \citep[cf.][]{savitsky2022}. 
$$
\boldsymbol h_t = (\cos(2\pi\omega_{k_1}t), \sin(2\pi\omega_{k_1}t), \ldots, \cos(2\pi\omega_{K}t), \sin(2\pi\omega_{K}t))^\top
$$
where the term $\omega_k=k/T$ is a Fourier frequency for which the associated sinusoid completes an integer number of cycles in the observed length of time series, and $K$ cannot be larger than $T/2$. % Which frequencies shall be included in the covariate matrix can be decided by inspecting the periodograms of the time series for each location~$i$ (see Figure~\ref{fig:agg_grid}).
In order to choose the dimension of the harmonic regressions, we inspect the estimated spectral densities of our $I=21\times 21=441$ time series (see Figure~\ref{fig:ts_spectral}) and observe the strong intra-daily 
as well as an intra-weekly seasonality, with the largest values corresponding to 24-hour intervals. We use this information to select the 12 frequencies marked by vertical
lines in Figure~\ref{fig:ts_spectral} in constructing the harmonic regression. This means
that the dimension of our vector of covariates $\boldsymbol h_t$ is 24. 

Moreover, in order to capture the spatio-temporal dependence in crowdedness, the error term $\eta_{i,t}$ is split into two components:
$$
\eta_{i,t} = w_{i,t} + e_{i,t},
$$
where $e_{i,t}$ is normally distributed $e_{i,t} \stackrel{iid}\sim N (0, \sigma^2)$ and $wi,t$ is a space-time random
effect which captures the spatio-temporal dependence in the log crowdedness measure
unexplained by the Fourier covariates.

In the following, we investigate and compare the predictive performance of different
models for the random effects. 
As a baseline model, we consider a model with no random effects  $w_{i,t}=0$ and with one set of regression coefficients for all locations as the baseline. 
The most complex model we employ is one proposed in \cite{modelac}, where $w_{i,t}$ has a spatio-temporal specification and the area-specific coefficients $\bm\beta_i$
are clustered using a Bayesian non-parametric procedure (CAR-AR-BNP). This allows us 
to identify clusters of areas with similar temporal behavior while
also keeping the model more parsimonious.

In addition to the CAR-AR-BNP model, we consider three CAR-AR models which all assume a common vector or regression coefficients for all areas with a normal prior: 
i) a model where the spatial auto-correlation parameter $\rho$ is fixed to $0$ (CAR-AR ($\rho=0$)) -- this only implies a temporal dependence structure, 
ii) a model where spatial auto-correlation parameter $\rho$ is fixed to $0.5$  (CAR-AR ($\rho=0.5$)) -- this only implies temporal dependence structure and a moderate spatial dependence,
iii) a model where the spatial auto-correlation parameter $\rho$ is estimated in the MCMC procedure (CAR-AR). 
More details about the models and the estimation procedure using MCMC can be found in  Appendix~\ref{sec:model}.

To evaluate the performance of the different models, 
we set up an out-of-sample exercise based on rolling windows, where we start by training the Bayesian model on data between November 04, 2013, at 00:00 (Monday) and November 10, 2013, 23:50 to generate one-, two- and three-step-ahead predictions, as well as for computing the predictive measures to be used for model selection for November 11, 2013, 00:00 up to November 11, 2013, 00:20. 
In a separate estimation procedure, we shift the window of the training data by 10 minutes and re-estimate the model in an iterative fashion until we reach the end of the sample. 

All results are based on 10000 iterations of the Gibbs sampler, where the first 5000 are discarded as burn-in and the thinning parameter is set to 50. This leaves 100 draws to be used for inference. The number of draws is not as large as 
typical values, but we use it to keep the verification of the properties (especially 
for Property \ref{p4}) manageable in terms of computation time on a local computer. 
If one has access to a cluster of workstations, the number of iterations can be increased,
as the property verification can trivially be parallelized. The trace plots of the models show acceptable convergence for all parameters, as well as good mixing.

\begin{figure}[tp]
    \centering
    \includegraphics[width=\columnwidth]{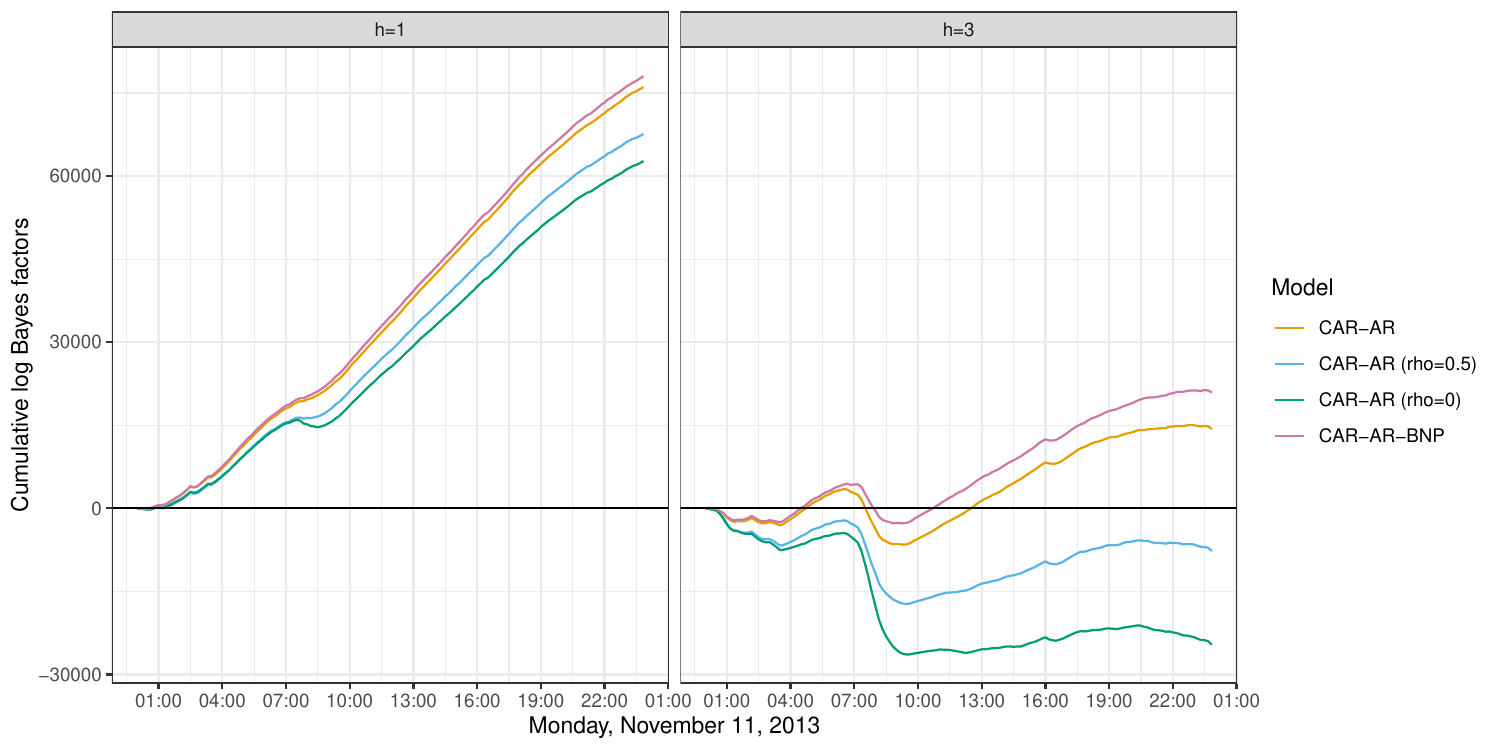}
    \caption{One- and three-steps-ahead cumulative log predictive Bayes factors for different models relative to the baseline model.}
    \label{fig:cum_log_bf}
\end{figure}

Figure~\ref{fig:cum_log_bf} presents the cumulative one-step ahead ($h=1$) and three-step ahead ($h=3$) log predictive Bayes factors for Monday, November 11, 2013:
$$
\log\text{BF}_{t_1,t_2}(A,B)=\sum_{t=t_1}^{t_2-h}\log\text{LPDS}_{t+h}(A)-\log\text{LPDS}_{t+h}(B),$$
where $B$ is taken to be a baseline model.
We observe that all models taking space and/or time correlation into account through the autoregressive structure outperform the baseline model, with the model  CAR-AR-BNP with spatial clustering also outperforming, even if not by a lot, the CAR-AR model. 
In terms of the three-steps ahead prediction, the CAR-AR-BNP model is superior, but we observe also that the performance of all CAR-AR models is worse than the baseline in the hours following midnight and between 07:00-09:00 (see negative slopes in the log Bayes factor curves). Furthermore, the gain in performance of the CAR-AR models diminishes around 19:00, when the slopes of the curves are not as steep.
These results point towards the fact that there might be a change in the spatial dependence parameter $\rho$ throughout the day. We leave such an extension of the model to further research.

\begin{table}[tp]
\caption{Posterior means and standard deviations (in parentheses) of the predictive measures of accuracy and F1 scores for property satisfaction and the RMSE of the robustness for the four properties. The reported measures are averages over all test samples.}
\centering
\begin{tabular}{lccccc}
  \toprule
  Measure&Baseline& CAR-AR&CAR-AR&CAR-AR&CAR-AR  \\ 
   &&$\rho=0$&$\rho=0.5$&&  BNP\\ 
  \midrule \multicolumn{6}{c}{Property \ref{p1}} \\ \midrule
$\bar{\text{Acc}}^{\text{Satisf}}$ & 0.6999 & 0.944 & 0.9479 & 0.9445 & 0.9498 \\ 
   & (0.0056) & (0.0025) & (0.003) & (0.0059) & (0.0035) \\ 
  $\bar{\text{F1}}^{\text{Satisf}}$ & 0.8027 & 0.9555 & 0.9584 & 0.956 & 0.9601 \\ 
   & (0.0029) & (0.0019) & (0.0022) & (0.0043) & (0.0026) \\ 
  $\text{RMSE}^{\text{Rob}}$ & 493.2178 & 101.979 & 96.0108 & 101.7701 & 92.9006 \\ 
   & (5.84) & (3.8052) & (4.2172) & (9.328) & (6.104) \\ 
  \midrule \multicolumn{6}{c}{Property \ref{p2}} \\ \midrule
$\bar{\text{Acc}}^{\text{Satisf}}$ & 0.8177 & 0.9463 & 0.9481 & 0.9476 & 0.9553 \\ 
   & (0.0056) & (0.0038) & (0.0038) & (0.0043) & (0.0037) \\ 
  $\bar{\text{F1}}^{\text{Satisf}}$ & 0.8988 & 0.9678 & 0.9688 & 0.9686 & 0.9731 \\ 
   & (0.0033) & (0.0022) & (0.0023) & (0.0025) & (0.0022) \\ 
  $\text{RMSE}^{\text{Rob}}$ & 273.7798 & 59.9263 & 57.5646 & 59.1099 & 48.6895 \\ 
   & (8.4465) & (2.771) & (2.5964) & (2.7259) & (2.8436) \\ 
  \midrule \multicolumn{6}{c}{Property \ref{p3}} \\ \midrule
$\bar{\text{Acc}}^{\text{Satisf}}$ & 0.9116 & 0.9511 & 0.9508 & 0.9515 & 0.9547 \\ 
   & (0.0076) & (0.0028) & (0.0031) & (0.0036) & (0.0039) \\ 
  $\bar{\text{F1}}^{\text{Satisf}}$ & 0.9439 & 0.9692 & 0.969 & 0.9695 & 0.9713 \\ 
   & (0.0051) & (0.0018) & (0.0021) & (0.0024) & (0.0026) \\ 
  $\text{RMSE}^{\text{Rob}}$ & 120.5187 & 59.3011 & 59.1511 & 58.645 & 52.6133 \\ 
   & (7.339) & (1.9052) & (2.8903) & (3.8598) & (3.461) \\ 
  \midrule \multicolumn{6}{c}{Property \ref{p4}} \\ \midrule
$\bar{\text{Acc}}^{\text{Satisf}}$ & 0.9268 & 0.9703 & 0.9722 & 0.9726 & 0.9743 \\ 
   & (0.0061) & (0.0034) & (0.0037) & (0.0043) & (0.0034) \\ 
  $\bar{\text{F1}}^{\text{Satisf}}$ & 0.8575 & 0.939 & 0.9426 & 0.9437 & 0.947 \\ 
   & (0.0117) & (0.007) & (0.0075) & (0.0084) & (0.0068) \\ 
  $\text{RMSE}^{\text{Rob}}$ & 280.6554 & 130.225 & 125.8994 & 128.1106 & 122.4351 \\
   & (20.4942) & (6.0651) & (5.6114) & (6.5048) & (5.1467) \\ 
  \bottomrule
\end{tabular}
\label{tab:formalpredmeasures}
\end{table}

\begin{figure}[tp]
    \centering
    \includegraphics[width=\columnwidth]{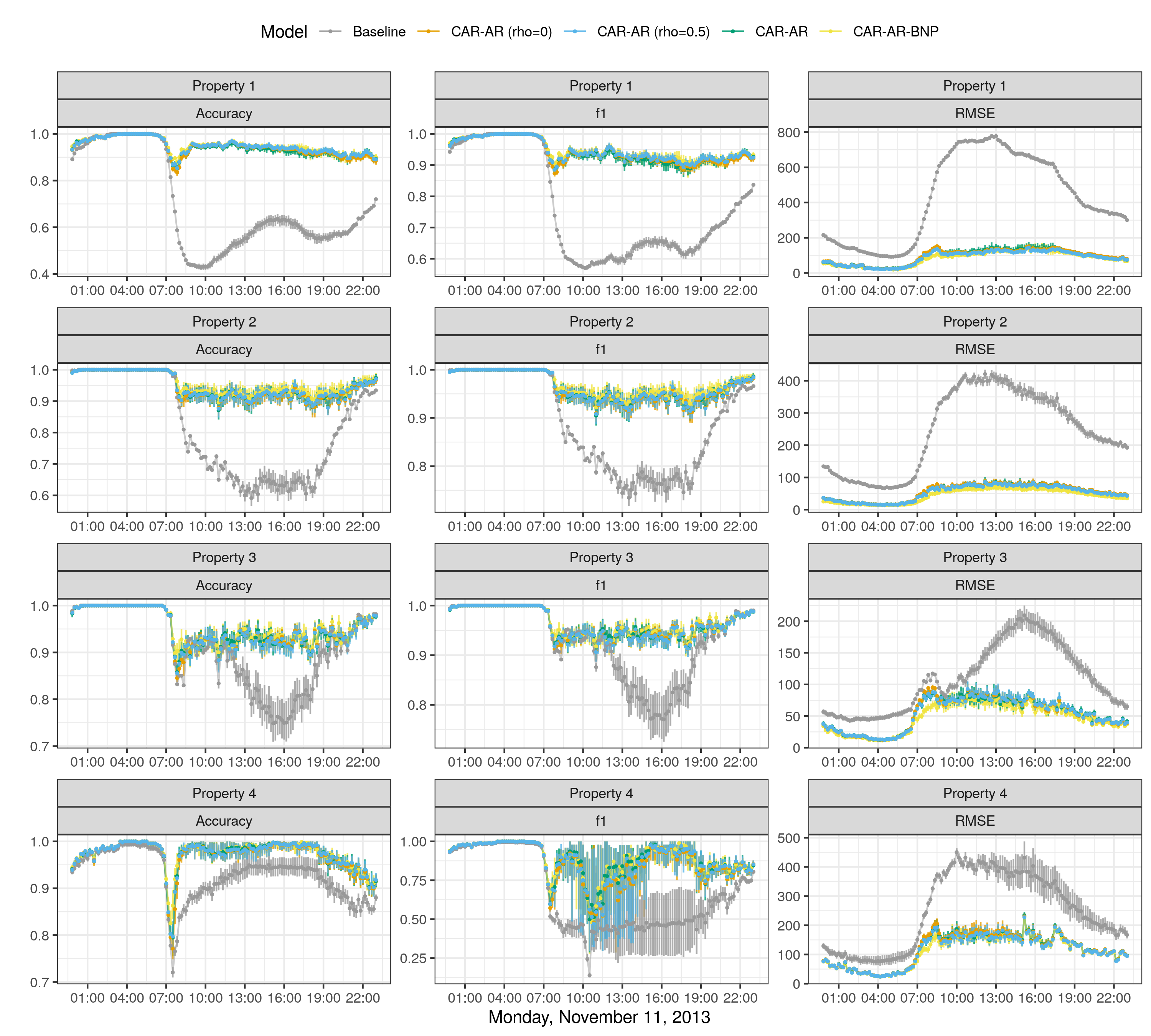}
    \caption{Mean accuracy, F1 score for the Bayesian predictive satisfaction measure and the RMSE of the robustness measure (points) with their corresponding 80\% credible intervals for a rolling window exercise.}% where the time point of the $x$-axis represents the end of the training sample. }
   \label{fig:formal_measures} 
\end{figure}

Finally, we also investigate the performance of the five models in terms of the predictive measures based on property satisfaction and robustness introduced in Section~\ref{sec:property_comparison}. The property parameters used for all properties are $c=500$, $h_\text{P.1}=h_\text{P.3}=30~\text{min}$, $h_\text{P.2}=10~\text{min}$,
$h_\text{P.4}=40~\text{min}$,
$d_\text{P.2}=d_\text{P.3}=1~\text{cell}$, $d_\text{P.4}=4~\text{cells}$.
Table~\ref{tab:formalpredmeasures} shows the posterior mean and standard deviation of the satisfaction accuracy, satisfaction F1 score and robustness RMSE for all four properties. We observe that the CAR-AR-BNP model is the best-performing one in terms of the measures inspected, however, the difference in performance for some properties is not large. 
Figure~\ref{fig:formal_measures} presents the average value of the measures in Table~\ref{tab:formalpredmeasures} for all testing periods, together with 80\% credible intervals.  This figure can be used for deciding which model performs best in terms of specific interest in the verified properties. For example, it can be seen that the autoregressive models perform similarly in terms of satisfaction measures for all properties, while the robustness of the model CAR-AR-BNP is better for properties \ref{p2} and \ref{p3}. The same model also outperforms the others in terms of property \ref{p4} during the rush hours 07:00-09:00, so it should be chosen if the performance in this specific time frame is of interest to the modeler.

\subsection{Results of the spatio-temporal model with spatial clustering}
%The results in this section are based on 10000 iterations of the Gibbs sampler, where the first 5000 are discarded as burn-in and the thinning parameter is set to 50.  The values of the hyper-parameters are chosen as $\bm\mu_{0}=\bm 0$, $\Sigma_{0} = 0.1 \mathbb{I}_{2K}$, which constitutes an informative prior with almost all prior mass on the interval ($-1, 1$); for the inverse gamma priors we choose $a^\sigma=a^\tau=1$ and  $b^\sigma=b^\tau=0.01$ which are weakly informative priors with a prior mean for the variance parameters of around 0.09. 
%For the auto-regressive parameters, we set $a^\xi=2$ and $b^\xi=1$ which imply a prior mean of around 0.3 and a prior mode of around 0.85.
%We take $\alpha=1$ in the BNP prior and we use 50 auxiliary variables in the algorithm employed for sampling the cluster assignments.

\begin{figure}[tp]
     \centering
         \includegraphics[width=0.6\textwidth]{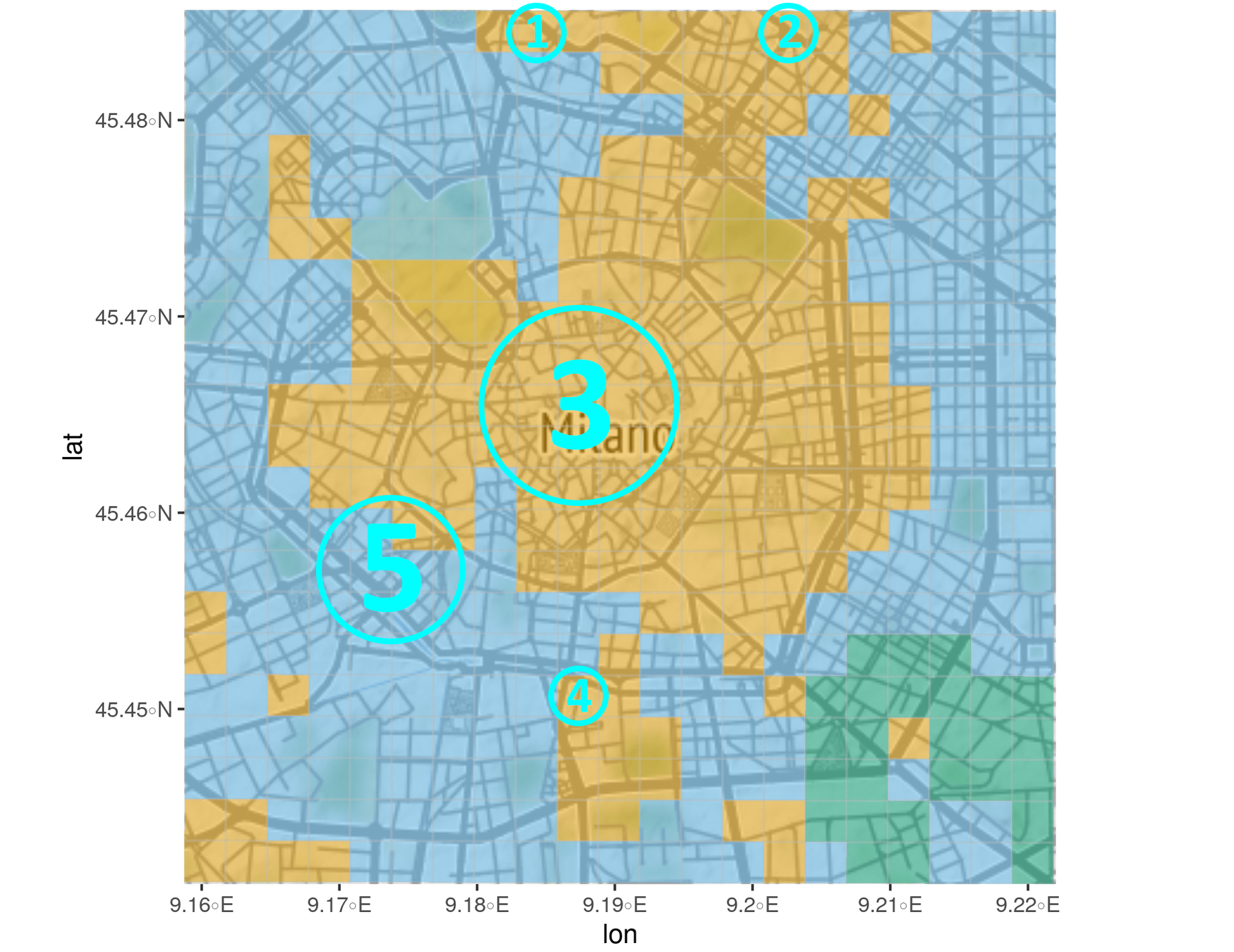}
         \includegraphics[width=0.8\textwidth]{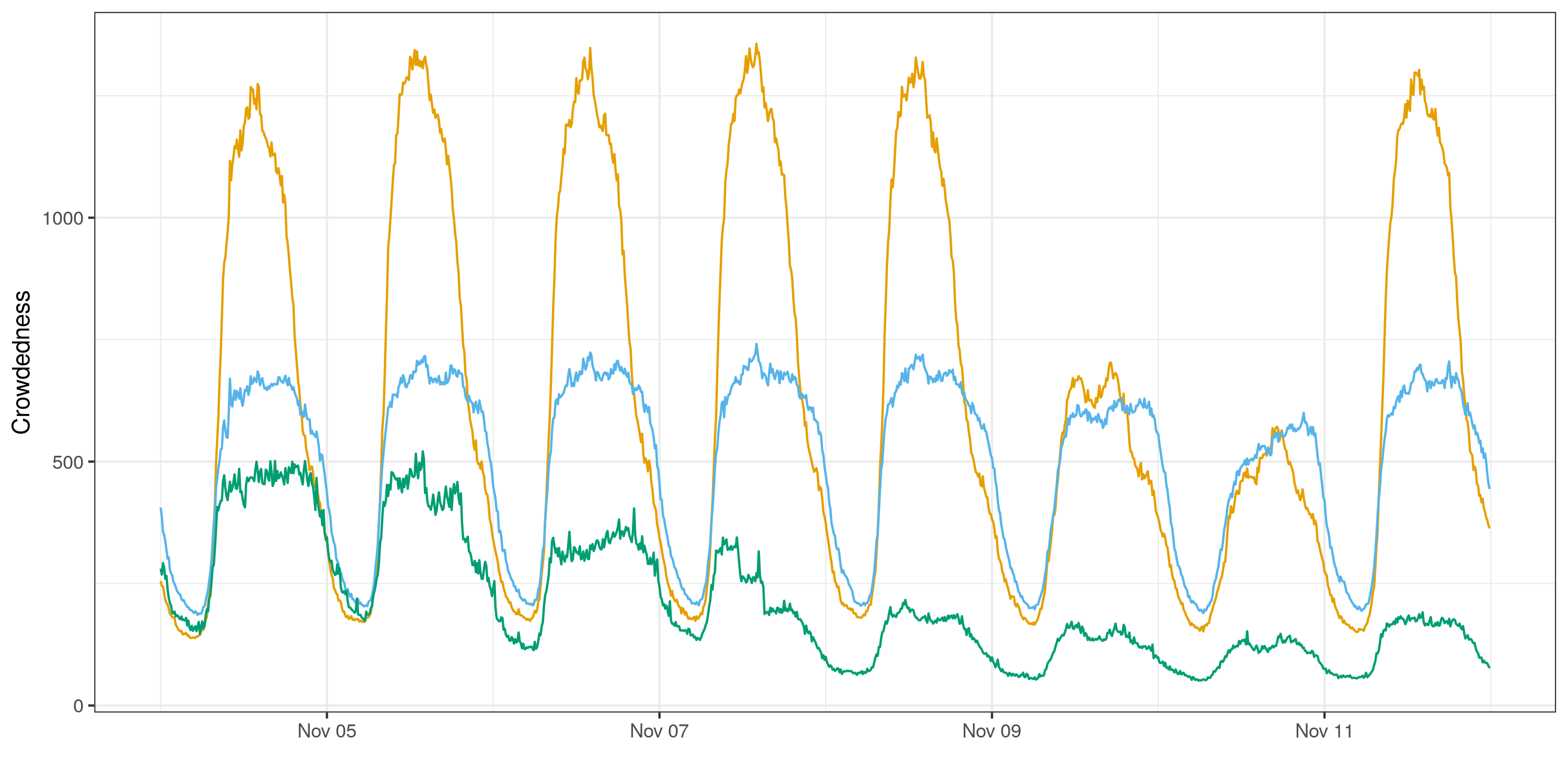}
        \caption{Top panel: Clusters as identified by Binder's loss. Bottom panel:
     crowdedness measure averaged over all locations in one cluster. The marked areas in the top panel represent: \circled{1}~\emph{Garibaldi Station}, \circled{2}~\emph{Central Station}, \circled{3}~\emph{Duomo}, \circled{4}~\emph{Bocconi}, \circled{5}~\emph{Navigli}.
     }
    \label{fig:map_ts_clusters}
\end{figure}

We present in the following the results of the CAR-AR-BNP model.
The top panel in Figure~\ref{fig:map_ts_clusters} presents the three clusters identified by employing Binder's loss on the samples of the cluster assignments vector, while the bottom panel presents the crowdedness measure averaged over all locations in one cluster. 
% cluster 1
The blue cluster is one where the difference in the crowdedness between weekends and working days is not as large as for the other two, with activity peaking in the morning (stronger during the working days) as well as in the evening (stronger effect on Sunday). The \circled{5}~\emph{Navigli} area is a member of the blue cluster. 
% cluster 2
The yellow cluster contains areas where the activity is high on the working days and lower on the weekends, with an intraday peak around noon. Typical locations in this cluster are university centers or the city center, where most office buildings are situated.
%% cluster 3
The green cluster is the smallest one, with the characteristic that the activity plummets during the weekend. The area corresponding to this cluster is Porta Romana, which contains the train station with the same name, a station primarily used by commuters into the city.
Moreover, the rather isolated areas belonging to one cluster but enclosed by areas in other clusters seem to be explainable and likely not caused by model artifacts. For example, the yellow square in the middle of the green cluster is the location of a large shopping mall.
%
% \begin{figure}[htbp!]
%     \centering
%     \includegraphics[width=0.9\textwidth]{Figures/clusters_21x21.png}
%      \includegraphics[width=0.8\textwidth]{Figures/y_mean_cluster.png}
%     \caption{Top panel: Clusters as identified by Binder's loss. The cells marked with borders represent the districts of Duomo (black), Bocconi (red) and Navigli (blue). Bottom panel:
%     crowdedness measure averaged over all locations in one cluster
%     %The posterior mean of the harmonic regression for the three clusters.
%     }
%     \label{fig:map_ts_clusters}
% \end{figure}
%

%
Finally, posterior means of the parameters that measure time and space dependence indicate strong persistence in both space and time.

\subsection{Verification of the crowdedness requirements for spatio-temporal model with spatial clustering}

\subsubsection{\ref{p1} -- Overloads are temporary}
\begin{figure}[tp]
    \centering
    \includegraphics[width=1\linewidth]{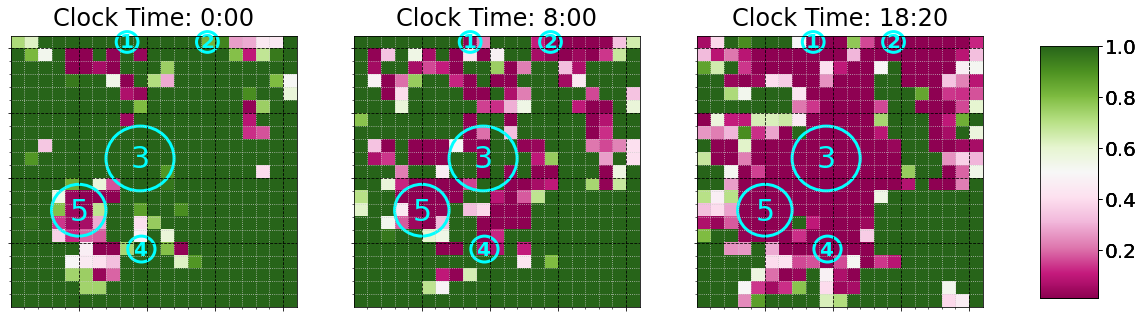}
    \includegraphics[width=1.0255\linewidth]{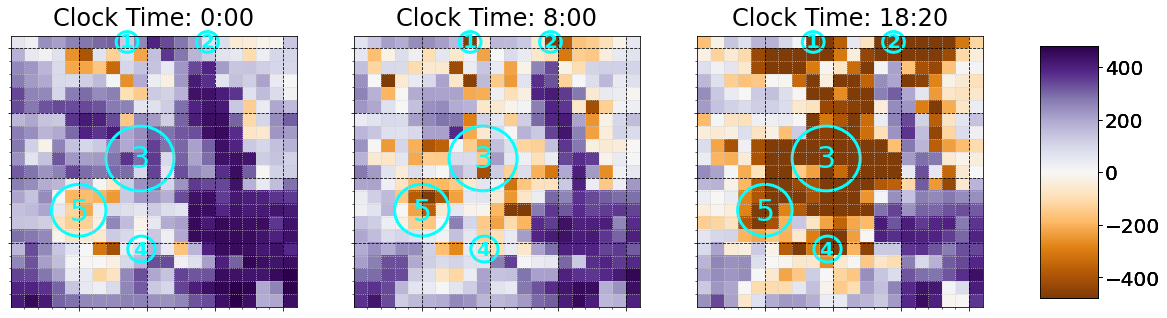}
    \caption{Posterior satisfaction probability  (top row) and posterior mean of the robustness measure (bottom row) resulting from checking for property~\ref{p1} at three times of the day.
    The marked areas represent: \circled{1}~\emph{Garibaldi Station}, \circled{2}~\emph{Central Station}, \circled{3}~\emph{Duomo}, \circled{4}~\emph{Bocconi}, \circled{5}~\emph{Navigli}.}
    \label{fig:p1-smc}
\end{figure}

Figure~\ref{fig:p1-smc} shows the estimated posterior satisfaction probability and the posterior mean of the robustness measure resulting from the evaluation of property~\ref{p1}.
Note that~\ref{p1} defines a property only in terms of a temporal operator, where we set $h_\text{P.1} = 30~\text{min}$. That is, we check whether the crowdedness variable stays below a value of $c=500$ or, in case it exceeds this value, then it must return below it within 30 minutes. When looking at the results, it appears clear that the city is roughly split into two macro-areas: the historical and financial center is unlikely to satisfy the property during busy times, while the residential areas are almost always satisfying it. A relevant exception comes from the \circled{5}~\emph{Navigli} area (left bottom of the map): it is, in fact, a vibrant area, where many young people live, which has many touristic landmarks and an active commercial area.
We can see that this area is consistently violating our requirement over time, although, from a look at the robustness its actual value is close to zero, meaning that the violation is quite small, making it less concerning from a network capacity perspective.
%We can see that, from a temporal perspective it is the area that is consistently violating the property we have defined, although, a look at the robustness shows that minimal interventions on the network can solve the issue, since the actual robustness values are near to zero for most of the day, meaning that the violation of the requirement is not very significant.
%
\subsubsection{\ref{p2} -- Overloads are local}

\begin{figure}[tp]
    \centering
    \includegraphics[width=\linewidth]{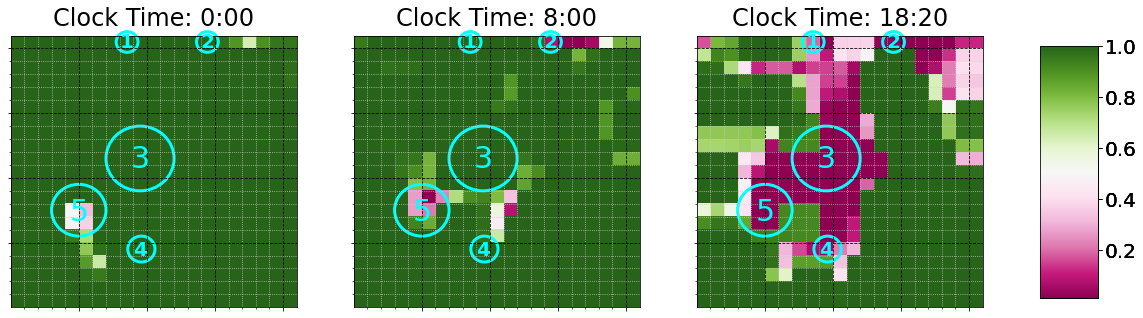}
    \includegraphics[width=1.0225\linewidth]{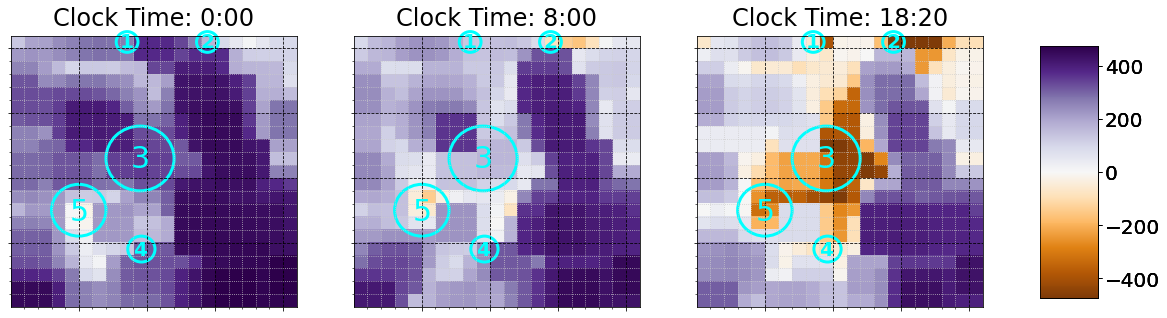}
    \caption{Posterior satisfaction probability  (top row) and posterior mean of the robustness measure (bottom row) resulting from checking property~\ref{p2} at three times of the day.
    The marked areas represent: \circled{1}~\emph{Garibaldi Station}, \circled{2}~\emph{Central Station}, \circled{3}~\emph{Duomo}, \circled{4}~\emph{Bocconi}, \circled{5}~\emph{Navigli}.}
    \label{fig:p2-smc}
\end{figure}

Figure~\ref{fig:p2-smc} shows the estimated posterior satisfaction probability and the posterior mean of the robustness measure resulting from the evaluation of property~\ref{p2} at three different times of the day.
Note that this property is based only on predictions
%only exploits spatial operators, so its violation is only dependent on the values 
of neighboring locations ($d_\text{P.2} = 1~\text{cell}$) at future time $t+h_\text{P.2}$, with $h_\text{P.2} = 10~\text{min}$. %We consider here only the 10 min-ahead predictions. %\circled{5}
%The ``somewhere'' operator in \ref{p2} basically imposes that each location is connected to another one exhibiting different values of crowdedness. 
%As one might expect, the property is likely satisfied practically everywhere for most of the day, while the only areas that are violating it more frequently are the \circled{3}~Duomo ones and the ones moving from them. 
As one might intuitively expect, the property exhibits high values of satisfaction and robustness for a large area of the city center (there is usually at least an uncrowded area connected to a crowded one). A notable exception is the \circled{3}~\emph{Duomo} area, from which crowds spread towards the other hot spots at the busiest time of the day (18:20). 
%In that case, a look at the robustness shows that the \circled{3}~Duomo area reaches the lowest robustness levels, together with the \circled{2}~Central Station.
However, by looking at the posterior predictive mean of the robustness for different time points, we get a clearer understanding of the spatial distribution of the excessive loads. In fact, it is evident that the \circled{3}~\emph{Duomo} is the area that might most likely suffer from excessive crowdedness, without any possibility of enacting load-balancing strategies based on the state of nearby locations.
%With this regard, it is interesting to look at the satisfaction probability: isolated locations having high satisfaction probability consistently over time can be alert for some pathological behaviors of the system (for instance, they might mark locations where the value of crowdedness is misreported or locations that are physically isolated for nearby ones and so that there are wasted networking resources in place). A look at the average robustness, on the other hand, can help in understanding whether such 
Conversely, other areas, like the \circled{5}~\emph{Navigli} area, have a much safer spatial behavior, either because they exceed the threshold only slightly, 
% is almost always slightly above the threshold, 
or because they are surrounded by areas with much lower levels of crowdedness.
%of disproportionate levels of crowdedness in comparison to nearby areas, or simply because they happen to be very close to the threshold, but always slightly above, while the nearby locations are always slightly below.  

\subsubsection{\ref{p3} -- The network is fault-tolerant}

\begin{figure}[tp]
    \centering
    \includegraphics[width=\linewidth]{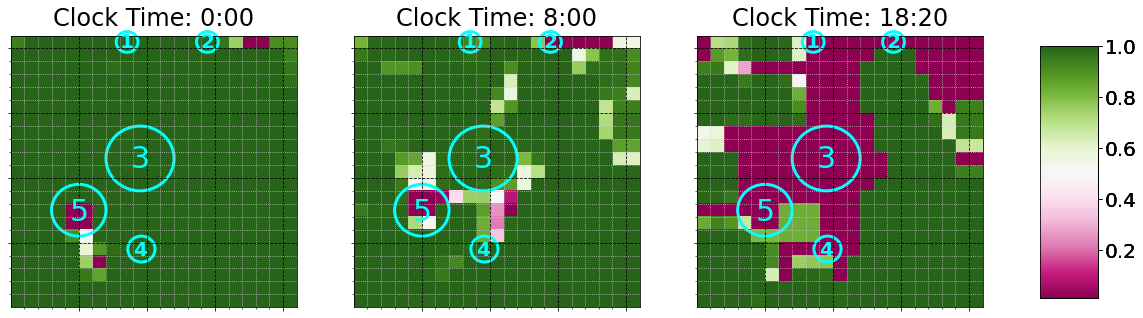}
    \includegraphics[width=1.0225\linewidth]{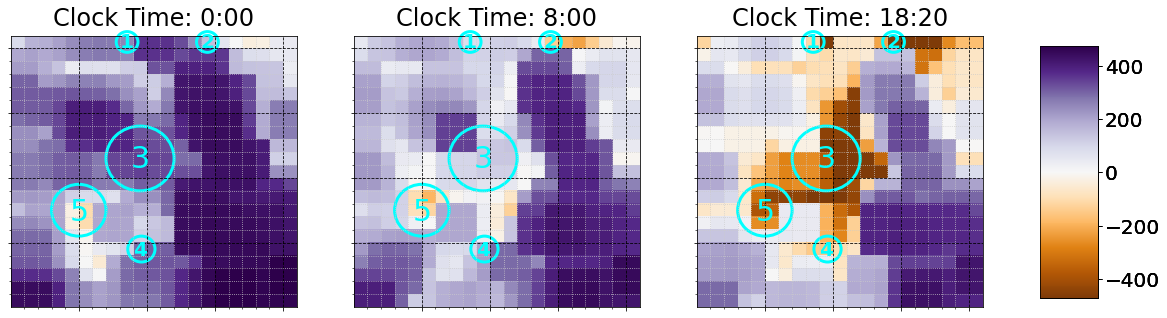}
    \caption{Posterior satisfaction probability  (top row) and posterior mean of the robustness measure (bottom row) resulting from checking property~\ref{p3} at three times of the day. The marked areas represent: \circled{1}~\emph{Garibaldi Station}, \circled{2}~\emph{Central Station}, \circled{3}~\emph{Duomo}, \circled{4}~\emph{Bocconi}, \circled{5}~\emph{Navigli}.}
    \label{fig:p3-smc}
\end{figure}

Figure~\ref{fig:p3-smc}  shows the estimated posterior satisfaction probability and the posterior mean of the robustness measure resulting from the evaluation of property~\ref{p3} at three different times of the day.
Property \ref{p3} enforces the availability of a neighboring uncrowded area (i.e., $d_{\text{P.}3} = 1$) consistently for $h_{\text{P.}3} = 30~\text{min}$.
% might be a little harder to grasp, because of the need to take into account both nearby locations and future values at the same time.
%temporal and spatial effects coming from the in contrast with previous properties define a spatio-temporal requirement, i.e. a property that must hold in different areas at each time of the day, depending on current and future values of neighboring locations.
The first thing the reader might notice is that this property exhibits a visual pattern similar to \ref{p1}--\ref{p2}, except that it is in general less likely to be satisfied.
% If one confronts the picture precisely, it can be seen that for practically every location of \ref{p2}, the observed value of likelihood of satisfaction or average robustness is lower (more pink/orange areas and with stronger colors). % with an overall average robustness that is lower in almost every location. 
This behavior is to be expected, as one can notice when looking at the logic formulas: \ref{p3} resembles, in fact, the structure of the right side of implication (``$\rightarrow$'') in \ref{p1}--\ref{p2}, except that it enforces stricter requirements (there must be an uncrowded area for the next half-hour). This observation shows a key strength of logic for the validation and explainability of specifications: from an informal perspective, \ref{p1} and \ref{p2} describe different aspects than \ref{p3}. Yet, the obtained results show that \ref{p3} could effectively replace \ref{p1}--\ref{p2} as a specification that encompasses both of them.
% That is because the property can be partly seen as a combination of the previous ones, and in fact, the area with a satisfaction probability higher than 0.5 closely maps the intersection of the satisfaction probability areas of the previous properties. 
%The left-hand side of the implication ``$\rightarrow$'', in fact, is similar to the one of \ref{p1} and \ref{p2}, except that it is slightly weaker, in the sense that it is not be satisfied only in cases of locations having a high level of crowdedness consistently for at least thirty minutes. The right-hand side of the ``$\rightarrow$'', on the other hand, combines a lighter version of \ref{p2} with a stricter version of the right-hand side of $\ref{p1}$, as it expresses that during that time in which the location is exhibiting an overload, there must be an adjacent location not saturated so that a proper load-balancing routine can be triggered to avoid service reductions and system failures. 
In fact, \ref{p3} summarizes the ideal behavior of a fault-tolerant system overall. Verifying this property clearly shows that the city is split into two parts, with the most-touristic part less likely to satisfy the property, while the residential and non-touristic areas are more likely to satisfy it.
\subsubsection{\ref{p4} -- Uncrowded reachability}

\begin{figure}[tp]
    \centering
    \includegraphics[width=\linewidth]{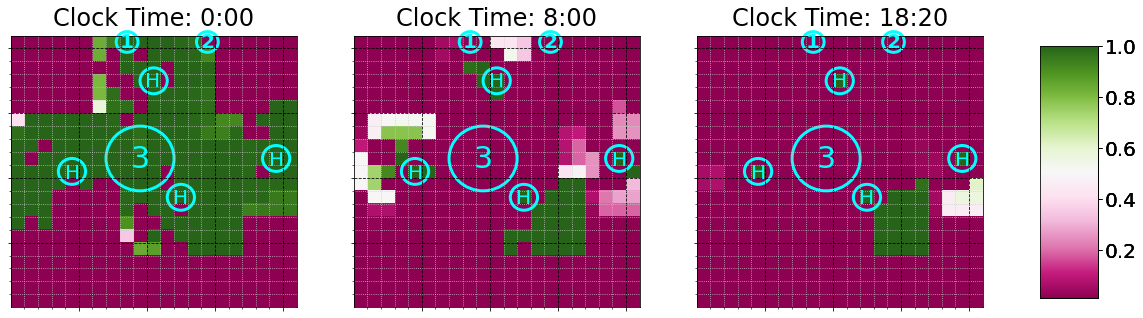}
    \includegraphics[width=1.0225\linewidth]{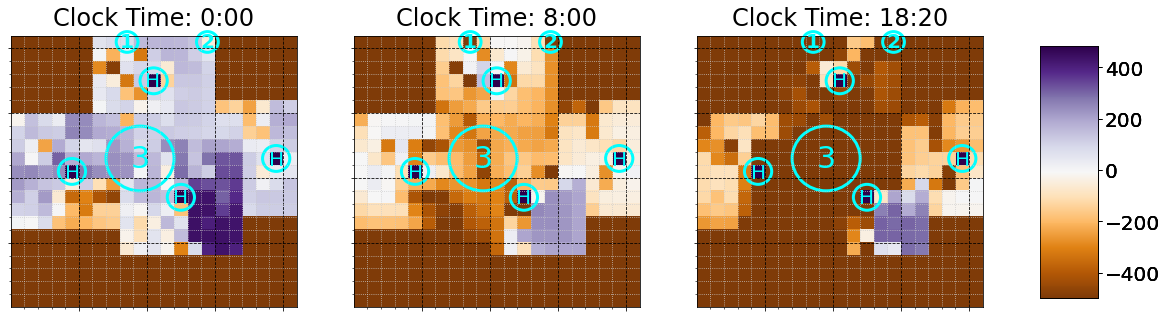}
    \caption{Posterior satisfaction probability  (top row) and posterior mean of the robustness measure (bottom row) resulting from monitoring the property~\ref{p4} at three times of the day. The marked areas represent: \circled{1}~\emph{Garibaldi Station}, \circled{2}~\emph{Central Station}, \circled{3}~\emph{Duomo}.}
    \label{fig:poi-smc}
\end{figure}

Figure~\ref{fig:poi-smc} presents the results of verifying property~\ref{p4} for $h_\text{P.4} = 40 \text{min}$ and $d_\text{P.4} = 4~\text{cells}$ (approximately $1$ km).
\circled{H} circles mark hospitals' locations, where the property is trivially satisfied in all circumstances.
By looking at the picture, an immediate observation is that areas at the corners are simply too far from any of the city center hospitals, meaning that going towards the center from there would be impractical. However, it is interesting to see that, while it is never very easy to get to hospitals in busy times (like at 18:20), the \circled{2}~\emph{Central Station} is still in a good spot (as it is not too far, and not too crowded), conversely the \circled{1}~\emph{Garibaldi Station} is in a less favorable location, as it becomes practically inaccessible in crowded times. Even worse is the \circled{3}~\emph{Duomo} area, which, despite being quite close to a hospital, experiences such high levels of crowdedness that make reaching the hospital almost impossible in crowded times (in terms of the requirement we have defined), while it is relatively easier in medium-crowded times. Lastly, the always failing areas at the corners of the grid, and at the bottom-center tell us something different: for the spatial configuration we are considering, they always violate the requirement to reach a hospital in the city center in $d_{\text{P.}4}$. This can be surprising at first, but a look at the broader map of the city clarifies that they are closer to hospitals that are not in our grid and, therefore cannot be fully analyzed by our model.

%%%%%%%%%%%%%%%%%%%%%%%%%%%%%%%%%%%%%%%%%%%%%%
\section{Discussion and future work}
\label{sec:conclusion}

In this paper, we propose a framework for predictive model checking and comparison, where in addition to usual approaches, we advocate for the specification of concrete (spatio-temporal) properties that the predictions from a model should satisfy. Given trajectories from the Bayesian predictive distribution, the posterior predictive probability of satisfaction and the posterior predictive robustness of these properties can be approximated by verifying the properties on each of the trajectories efficiently using techniques from formal verification methods. Finally, we can evaluate ex post the model by comparing the resulting measures with the values in the observed data.

We illustrate the approach by building a Bayesian spatio-temporal model for areal crowdedness extracted from aggregated mobile phone data in the city of Milan and by formulating properties that the crowdedness level in the city network should satisfy in order to robustly withstand critical events. 
We compare various model specifications which include a 
harmonic regression with and without random effects,
as well as a model which performs clustering of the area-specific harmonic regression coefficients. The model that performs clustering is indeed the one that 
performs best in terms of the proposed measures. This model can however be further refined to clustering over time or a temporal evolution of the persistence parameter for the autoregressive random effects.

The proposed framework advocates for exploiting the rich information that Bayesian predictive inference offers in the form of draws from the posterior predictive distribution of future values, by evaluating models also based on properties that can be directly translated into decision-making. Therefore, we showcase how different model specifications are then evaluated based on well-known performance measures but also on posterior predictive measures employed in formal verification, such as the satisfaction probabilities or the robustness measure.  

On a larger scale, by exploiting the synergy between Bayesian modeling and formal verification methods, we also advocate for the development and use of explainable algorithms where properties relevant to decision-making are incorporated into the data analytic process flow. 
Therefore, the proposed approach has a clear potential in the area of sustainable cities and urban mobility, as these applications deal with complex systems, with a multitude of stakeholders and with a pressing need for transparency in the decision-making process. We hope for the illustration in the current paper to open the way to further applications.

%%%%%%%%%%%%%%%%%%%%%%%%%%%%%%%%%%%%%%%%%%%%%%
%% Supplementary Material, if any, should   %%
%% be provided in {supplement} environment  %%
%% with title and short description.        %%
%%%%%%%%%%%%%%%%%%%%%%%%%%%%%%%%%%%%%%%%%%%%%%
% \begin{supplement}
% \stitle{Title of Supplement A}
% \sdescription{Short description of Supplement A.}
% \end{supplement}
% \begin{supplement}
% \stitle{Title of Supplement B}
% \sdescription{Short description of Supplement B.}
% \end{supplement}

\paragraph{Computational details and replication materials}
The computations have been performed on 25 IBM dx360M3
nodes within a cluster of workstations. Instruction for downloading the data set as well as an extensive description can be found in \cite{barlacchi2015multi}. The estimation of the Bayesian models can be performed using code in the repository at
%\url{https://mega.nz/file/0EBXHYaJ#us5KMG1vxd3d3hKs_QnssU8-K9pXspg9PG9bwt-mYxs}
\url{https://github.com/lauravana/CARBayesSTBNP}. The Moonlight tool is available at \url{https://github.com/moonlightsuite/moonlight}, while the specific problem instance related to this project, together with the data and the scripts for generating the figures, are available at
%\url{https://mega.nz/file/kAg2zCrL#tBPNzL09VswrUONPWmwVFvQTnW0mUkqWTNAiTWNoVo4}
\url{https://github.com/ennioVisco/bayesformal}.

%% ** The bibliograhy **
\bibliographystyle{ACM-Reference-Format}
\bibliography{refs.bib}

% ** Acknowledgements **
\begin{acks}[Acknowledgments]
The authors acknowledge funding from the Austrian Science Fund (FWF) for the project ``High-dimensional statistical learning: New methods to advance economic and sustainability policies'' (ZK 35), jointly carried out by the University of Klagenfurt, the University of Salzburg, TU Wien, and the Austrian Institute of Economic Research (WIFO).
\end{acks}

\clearpage
\begin{appendices}
%%%%%%%%%%%%%%%%%%%%%%%%%%%%%%%%%%%%%%%%%%%%%%
\section{Models}
\label{sec:model}

\paragraph{Baseline model}
The simplest model considered is a harmonic regression, where we assume that the dependence in crowdedness is explained by the harmonic regressors and with error term $\eta_{i,t}\stackrel{iid}{\sim} \mathcal{N}(0, \sigma^2)$. 
Moreover, we assume that all spatial units share the same temporal behavior with $\boldsymbol{\beta}_i\equiv \boldsymbol\beta$.

\paragraph{CAR-AR models with common harmonic regression coefficients}
The error term $\eta_{i,t}$ is split into two components:
$$
\eta_{i,t} = w_{i,t} + \epsilon_{i,t},
$$
where $\epsilon_{i,t}$ is normally distributed $\epsilon_{i,t}\stackrel{iid}{\sim} \mathcal{N}(0, \sigma^2)$ and 
$w_{i,t}$ is a space-time random effect that captures the spatio-temporal dependence in the log crowdedness measure unexplained by the Fourier covariates. The random effect $w_{i,t}$ is modeled as a stationary
first-order autoregressive process:
\begin{equation}\label{eq:ranef}
w_{i,t} =  \xi w_{i,t-1} + \sqrt{1-\xi^2} u_{i,t}, \qquad t=2,\ldots, T,    
\end{equation}
where  $\xi \in (-1,1)$ to ensure stationarity of the model and $u_{i,t}$ is a mean zero stationary spatial innovation process with variance $\tau^2$ which is independent over time but correlated over the spatial units:
$$
\boldsymbol{u}_{t} \stackrel{iid}{\sim} \mathcal{N}(\boldsymbol 0, \tau^2 Q(\rho, W)^{-1}).
$$
For the first time point we have $\boldsymbol{w}_{1} \sim \mathcal{N}(\boldsymbol 0, \tau^2 Q(\rho, W)^{-1})$.
The matrix $Q(\rho, W)$ denotes the spatial precision proposed in \cite{leroux2000estimation}:
$$Q(\rho, W) = 
\rho(\text{diag}(W\boldsymbol 1) - W) + 
(1-\rho)\mathbb I,$$
where $\boldsymbol 1$ is the $I \times 1$ vector of ones, while $\mathbb I$ is the $I \times I$ identity matrix.
In this spatial prior, $0\leq \rho \leq 1$ provides a measure of spatial dependence while the spatial auto-correlation is controlled by the symmetric $I\times I$ adjacency matrix $W$, where $w_{kl}$ is equal to one if area~$k$ shares an edge or a vertex with area $l$ and zero otherwise (the so-called queen contiguity).
This mean-zero normal prior on the spatial innovations is referred to in the literature as a spatial conditionally autoregressive (CAR) prior.
We assume again one common set of regression coefficients $\boldsymbol{\beta}_i\equiv \boldsymbol\beta$.

\paragraph{CAR-AR model with spatial clustering (CAR-AR-BNP)}
We modify the model introduced above in order to identify areas with similar seasonality patterns. We place a Bayesian non-parametric (BNP) Dirichlet process prior to the $\boldsymbol \beta_i$ coefficients for all locations: 
\begin{equation}\label{eq:bnpprior}
   \boldsymbol \beta_{i}|P \stackrel{iid}\sim P, \quad i=1,\ldots, I,\, \text{with}\,\, P \sim \text{DP}(\alpha, P^0),\quad P(\cdot)=\sum_{j=1}^\infty\pi_j\delta_{\boldsymbol\theta_{j}}(\cdot),
\end{equation}
where the random measure $P$ is represented as the infinite sum of the product of random weights $\pi_j$ and locations $\boldsymbol\theta_{j}\sim P^0$ and $\delta_\theta$ represents the point mass at $\theta$. The stick-breaking prior is assumed on the common weights $\pi_j/\prod_{i=1}^{j-1}(1-\pi_i)\sim \text{Beta}(1, \alpha)$
and the reference measure is specified as
$
P^0(\boldsymbol\beta)=
\mathcal{N}_{2K}(\boldsymbol\beta|\boldsymbol m_0, S_0)
%\times \text{Beta}_{(-1,1)}(\boldsymbol\xi|a^{\xi},b^{\xi}),
$.
\paragraph{Further priors and estimation}

The intercept term $\beta_0$ has a mean zero normal prior. For the models with one set of regression coefficients, we employ $\boldsymbol\beta\sim\mathcal{N}_{2K}(\boldsymbol\beta|\boldsymbol m_0, S_0)$. Uniform priors are set on $\rho$ and  $\xi$ and inverse Gamma conjugate priors are set for the variance parameters: $\sigma^2\sim \mathcal{IG}(a^\sigma, b^\sigma)$ and  $\tau^2\sim \mathcal{IG}(a^\tau, b^\tau)$. These component specifications, along with our apriori independence assumption, form the joint prior.

% \subsection{Estimation}\label{sec:estimation}
Inference is performed using MCMC methods. The CAR-AR models are Gaussian state space models where the full conditional distributions of the parameters have a closed form, so a Gibbs sampler can be employed.   A rough outline of the samplers is given below: 
\begin{enumerate}
\item For CAR-AR-BNP: The marginalized sampler together with the reuse algorithm in \cite{favaro2013mcmc} is implemented for sampling the cluster assignments and the unique values of the cluster parameters \citep[cf.][]{modelac, mozdzen2022}. 
\item The unique values of the regression coefficients are sampled from the full conditional.
\item To sample the spatio-temporal random effects $w_{i,t}$ efficiently from an $I\times T$ multivariate normal distribution, we exploit the sparsity of the spatial precision matrix \citep[cf.][]{knorr-held2002, mccausland2011simulation, banerjee2017}.
\item Parameters $\rho$ and $\xi$ are sampled from a truncated normal distribution on the intervals $[0,1]$ and $[-1,1]$, respectively.
\item Variance parameters $\tau^2$ and $\sigma^2$ are sampled from the respective conjugate full conditional posterior distribution.
\end{enumerate}

For all models, the values of the hyperparameters are kept identical: $\bm m_{0}=\bm 0$, $S_{0} = 0.1 \mathbb{I}_{2K}$, $a^\sigma=a^\tau=1$,  $b^\sigma=b^\tau=0.01$.
We take $\alpha=1$ in the BNP prior, and we use $C=50$ auxiliary variables in the algorithm for sampling the cluster assignments \citep[see Section~3.2.1 in][]{favaro2013mcmc}. 
%%%%%%%%%%%%%%%%%%%%%%%%%%%%%%%%%%%%%%%%%%%%%%
\end{appendices}

\end{document}